\documentclass[a4paper,12pt]{article}

\usepackage{jheppub} 
\usepackage{booktabs}
\usepackage{extarrows}
\usepackage{multirow}
\usepackage[compat=1.1.0]{tikz-feynhand}
\usepackage{slashed}
\usepackage[normalem]{ulem}
\usepackage{subfig}
\usepackage{enumitem}

\allowdisplaybreaks[1]    

\definecolor{lblue}{rgb}{0.0,0.4,0.8}
\definecolor{lgreen}{rgb}{0,191,255}
\definecolor{lyellow}{rgb}{255,165,0}
\definecolor{dgreen}{rgb}{0,128,0}

\newcommand{\Lag}{\mathcal{L}}
\newcommand{\mC}{{\mathcal C}}
\newcommand{\mO}{{\mathcal O}}
\newcommand{\mB}{\mathcal{B}}
\newcommand{\mL}{\mathcal{L}}
\newcommand{\SM}{\text{SM}}
\newcommand{\NP}{\text{NP}}

\newcommand{\TeV}{\,\mathrm{TeV}}
\newcommand{\GeV}{\,\mathrm{GeV}}
\newcommand{\pb}{\,\mathrm{pb}}
\newcommand{\fbi}{\,\mathrm{fb}^{-1}}

\newcommand{\ymmp}{{y^{\mu\mu}_P}}
\newcommand{\ymms}{{y^{\mu\mu}_S}}

\newcommand{\ybbp}{{y^{bb}_P}}
\newcommand{\ybbs}{{y^{bb}_S}}
\newcommand{\yctp}{{y^{ct}_P}}
\newcommand{\ycts}{{y^{ct}_S}}
\newcommand{\yctl}{{y^{ct}_L}}
\newcommand{\yctr}{{y^{ct}_R}}

\newcommand{\Bsmumu}{B_s \to \mu^+\mu^-}

\newcommand{\ADG}{{\mathcal{A}_{\Delta\Gamma_s}^{\mu\mu}}}
\newcommand{\Smumu}{{\mathcal{S}_{\mu\mu}}}
\newcommand{\Cmumu}{{\mathcal{C}_{\mu\mu}}}
\newcommand{\Cmumulambda}{{\mathcal{C}_{\mu\mu}^\lambda}}
\newcommand{\BR}{\mB(\Bsmumu)}

\newcommand{\scenario}[2]{#1#2}

\subheader{
\hfill}

\title{Phenomenological anatomy of top-quark FCNCs induced by a light scalar singlet}

\author[a]{Biao-Feng Hou,}
\emailAdd{resonhou@zknu.edu.cn}

\author[a, b]{Xin-Qiang Li,}
\emailAdd{xqli@mail.ccnu.edu.cn}

\author[a, c]{Ya-Dong Yang,}
\emailAdd{yangyd@ccnu.edu.cn}

\author[a]{Xing-Bo Yuan}
\emailAdd{y@ccnu.edu.cn}

\author[a]{and Ming-Wang Zhang}
\emailAdd{mwzhang@mails.ccnu.edu.cn}

\affiliation[a]{Institute of Particle Physics and Key Laboratory of Quark and Lepton Physics (MOE),  Central China Normal University, Wuhan, Hubei 430079, China}
\affiliation[b]{Center for High Energy Physics, Peking University, Beijing 100871, China}
\affiliation[c]{Institute of Particle and Nuclear Physics, Henan Normal University, Xinxiang 453007, China}

\abstract{Scalar singlets under the Standard Model gauge group appear naturally in many well-motivated New Physics scenarios, such as the composite Higgs models. Unlike the Higgs boson in the Standard Model, they can induce large flavour-changing neutral currents (FCNCs) in the top sector. We investigate systematically the effects of a light scalar singlet $S$ with top-quark FCNC couplings, by including the low-energy constraints from the $B_s \to \mu^+ \mu^-$ decay, the muon anomalous magnetic moment $(g-2)_\mu$ and the neutron Electric Dipole Moment (EDM). We also perform a detailed Monte-Carlo simulation of the channel $pp \to t S +j$ with $S \to \mu^+ \mu^-$ and $S \to b \bar b$, and investigate the LHC sensitivity to the $tcS$ couplings. It is found that the scalar singlet $S$ can induce scalar-type contributions to the $B_s \to \mu^+ \mu^-$ decay, which do not suffer from the helicity suppression and contain a large CKM factor $V_{cs}^*V_{tb}$. As a result, constraints on the $tcS$ couplings from the measured branching ratio $\mathcal{B}(B_s \to \mu^+ \mu^-)$ are quite stringent, being even stronger than the expected LHC sensitivity in some parameter spaces. Besides the CP-conserving $tcS$ couplings, we have also considered the case of CP-violating $tcS$ couplings, with $y_{R,L}^{ct}=|y_{R,L}^{ct}|e^{i\theta_{R,L}}$. It is found that the CP observables $\mathcal{A}_{\Delta\Gamma_s}^{\mu\mu}$ and $\mathcal{S}_{\mu\mu}$ of the $B_s \to \mu^+ \mu^-$ decay are sensitive to the phase $\theta_R$, while the neutron EDM can provide bounds on the phase difference $\theta_L-\theta_R$. Therefore, they are complementary to each other in probing the CP phases of the $tcS$ couplings.}

\begin{document} 
\maketitle


\section{Introduction}
\label{sec:intro}

One major focus of the Large Hadron Collider (LHC) and its high-luminosity upgrade is the direct searches for new particles beyond the Standard Model (SM)~\cite{ZurbanoFernandez:2020cco}. While no credible signals have been found at the LHC so far, these searches are placing significant constraints on many New Physics (NP) models~\cite{ParticleDataGroup:2024cfk}. For many NP resonances, the current exclusion limits at the LHC have reached up to TeV or even multi-TeV scale~\cite{ExoticStatus}, e.g., lower limits for some particular types of $Z^\prime$ particles range from around $4$ to $5\TeV$~\cite{ATLAS:2019erb,CMS:2021ctt}. However, most of these limits rely on the couplings of the resonances to the first two generations, or assume flavour universal interactions. Alternatively, searches for new particles in top-quark related processes are also well motivated, since its large mass makes the top quark connect inherently to the origin of electroweak (EW) symmetry breaking~\cite{Beneke:2000hk,Schwienhorst:2022yqu,Franceschini:2023nlp}.

At the LHC, top quarks are copiously produced and searches for new particles in its rare decays hold, therefore, great promise. In this respect, the top-quark flavour-changing neutral current (FCNC) decays involving a new light scalar singlet ($S$) are regarded as an excellent probe of NP effects~\cite{Bahl:2023xkw,Castro:2022qkg}. The light scalar singlets are naturally present in many NP scenarios, including the Next-to-Minimal Supersymmetric Standard Model~\cite{Ellwanger:2009dp}, the Composite Higgs Models~(CHM)~\cite{Dimopoulos:1981xc,Kaplan:1983fs,Kaplan:1983sm,Cacciapaglia:2019bqz}, the Froggatt-Nielsen mechanism of flavour structures~\cite{Froggatt:1978nt}, and the EW Baryogenesis~\cite{Morrissey:2012db,Trodden:1998ym}. The top-quark FCNC couplings mediated by the scalar singlet $S$ may not be suppressed by the Glashow-Iliopoulos-Maiani mechanism, and thus significantly larger than those induced by the SM Higgs boson~\cite{ATLAS:2024mih,CMS:2024ubt}, as noticed in the CHM~\cite{Banerjee:2018fsx,Castro:2020sba}. From the viewpoint of Effective Field Theory~(EFT), the $S$-mediated FCNCs arise from dim-5 operators and thus less suppressed by one power of $v/\Lambda$ compared to those induced by the SM Higgs boson, which arise firstly at dim 6. Moreover, unlike the SM Higgs boson, the scalar singlet $S$ can decay into clean final states with relatively large branching ratios, such as $S\to \mu^+ \mu^-$  and $S\to b \bar b$. In light of these considerations, search strategies for such scalar singlets have been investigated at the LHC through the channels $p p \to t S + j$, with $S \to \mu^+ \mu^-$ or $S \to b \bar b$~\cite{Banerjee:2018fsx,Castro:2020sba}. Recently, the ATLAS collaboration has also performed a search for the scalar singlet in top-quark FCNC decays $t \to q S$, with $q=u,c$ and $S \to b \bar b$~\cite{ATLAS:2023mcc}. The obtained upper bounds on the branching ratio $\mB(t \to c S)$ are between $1.8\times 10^{-4}$ and $7.8 \times 10^{-4}$ for the scalar masses between $20$ and $160\GeV$.

The top-quark FCNCs mediated by new particles could also affect low-energy processes via the top-quark loops, such as the $B$-meson FCNC decays. In this respect, the $\Bsmumu$ decay is one of the most promising processes, which has already played an important role in constraining physics beyond the SM~\cite{Buras:2012jb,Buras:2013rqa,Fleischer:2017ltw,Fleischer:2017yox,Altmannshofer:2017wqy}. Recently, significant progress has been made in measuring this rare decay by the ATLAS~\cite{ATLAS:2018cur}, CMS~\cite{CMS:2022mgd}, and LHCb~\cite{LHCb:2021awg,LHCb:2021vsc} collaborations at the LHC. Their measured time-integrated branching fractions read
\begin{align}\label{eq:exp_br_bsmumu}
\begin{aligned}
  \BR_{\rm{ATLAS}} &= \left( 2.8^{+0.8}_{-0.7} \right)\times 10^{-9}, \\[0.2cm]
  \BR_{\rm{CMS}} &= \left( 3.83^{+0.38+0.19+0.14}_{-0.36-0.16-0.13} \right)\times 10^{-9},\\[0.2cm]
  \BR_{\rm{LHCb}} &= \left( 3.09^{+0.46+0.15}_{-0.43-0.11} \right)\times 10^{-9}. 
\end{aligned} 
\end{align}
Based on these results, the current Particle Data Group world average is given by~\cite{ParticleDataGroup:2024cfk}
\begin{align}\label{eq:br_bsmumu:exp}
  \BR_{\text{exp}} &= \left(3.34 \pm 0.27 \right)\times10^{-9}.
\end{align}
Within the SM, by including the QCD corrections up to next-to-next-to-leading order~(NNLO)~\cite{Buchalla:1993bv,Buchalla:1998ba,Misiak:1999yg,Hermann:2013kca} and the next-to-leading order~(NLO) EW correction~\cite{Bobeth:2013tba}, the branching ratio is predicted to be
\begin{align}\label{eq:br_bsmumu:SM}
  \BR_{\rm{SM}} = (3.51 \pm 0.09)\times10^{-9},
\end{align}
which is based on the calculations performed in refs.~\cite{Bobeth:2013uxa,Czaja:2024the} but with updated input parameters. For recent studies on the QED corrections, we refer to refs.~\cite{Beneke:2017vpq,Beneke:2019slt,Feldmann:2022ixt}. The above SM prediction is in good agreement with the current experimental average. Relevant studies in NP scenarios can be found, e.g., in refs.~\cite{Logan:2000iv,Buras:2013uqa,Li:2014fea,Chiang:2017etj,Arnan:2017lxi,Crivellin:2019dun,Lang:2022mxu}.

In this work, we will perform a systematic investigation of the top-quark FCNCs mediated by a new scalar singlet in the EFT framework, especially by including several low-energy constraints. Considering searches for such a singlet through the channel $pp \to t S + j$ with $S \to \mu^+ \mu^-$, the involved $tcS$ and $\mu\mu S$ couplings can also enter the $\Bsmumu$ decay at the one-loop level. It will be shown that this contribution results in the effective operators $(\bar s P_R b)(\bar \mu \mu)$ and $(\bar s P_R b)(\bar\mu \gamma_5 \mu)$ at the scale of $m_b$, with $m_b$ being the bottom-quark mass. Unlike the dominant SM contribution from $(\bar{s} \gamma^\mu P_{L} b)(\bar{\mu} \gamma_\mu\gamma_5\mu)$, these NP effects do not suffer from the helicity suppression. Furthermore, they are proportional to the CKM factor $V_{cs}^* V_{tb}$, which is much larger than $V_{ts}^* V_{tb}$ of the SM contribution. Therefore, it is crucial to study constraints on the scalar singlet $S$ from the $\Bsmumu$ decay. In order to constrain the $\mu\mu S$ coupling, bounds from the muon anomalous magnetic moment $(g-2)_\mu$ will also be considered. In addition, we will perform a detailed collider simulation, considering both the rare top decay $t \to q S$ and the single top production associated with an $S$ at the LHC, with the decay channels $S \to b \bar b$ and $S \to \mu^+ \mu^-$.

It is also noted that the $tqS$ couplings can be complex in general~\cite{Buras:2013rqa}, which is however difficult to access at the LHC by analyzing only the cleanest decay final states like $\mu^+ \mu^-$ and $b \bar b$. On the other side, besides the branching ratio, the $\Bsmumu$ decay offers other CP related observables, $\ADG$ and $\Smumu$, due to the sizable decay width difference of the $B_s$ system~\cite{DeBruyn:2012wk,Buras:2013uqa}. These CP observables are theoretically clean and can provide information on the CP-violating contributions from NP~\cite{Fleischer:2017yox,Fleischer:2024fkm}. Therefore, it is interesting to explore the potential of these CP observables in revealing the CP-violating $tqS$ interactions. In addition, the neutron Electric Dipole Moment (EDM) can generally provide extremely sensitive probes of new sources of CP violation beyond the SM~\cite{Pospelov:2005pr,Engel:2013lsa,Chupp:2017rkp,Alarcon:2022ero}. We will, therefore, investigate the neutron EDM constraint on the $tqS$ couplings and explore its complementarity with the $B_s \to \mu^+ \mu^-$ decay.

The article is organized as follows. In section~\ref{sec:EFT}, we introduce the EFT Lagrangian to describe the interactions between the scalar singlet $S$ and the SM fermions with top-quark FCNC couplings. In section~\ref{sec:low-energy constraints}, we recapitulate the theoretical frameworks for the $\Bsmumu$ decay, $(g-2)_\mu$ and the neutron EDM, and investigate constraints from these low-energy processes. In section~\ref{sec:collider}, we perform a detailed collider simulation and explore the phenomenology of $pp \to t S + j$ with $S \to \mu^+ \mu^-$ and $S \to b \bar b$ at the LHC. In section~\ref{sec:combined analysis}, we investigate the scalar singlet effects on the processes discussed in the last two sections, with both CP-conserving and CP-violating $tcS$ couplings. Our conclusions are finally made in section~\ref{sec:conclusions}. In appendix~\ref{sec:combined analysis:SP basis}, a combined analysis of the CP-violating $tcS$ couplings is performed in the scalar and pseudoscalar basis (i.e., in terms of the couplings $\ycts$ and $\yctp$), instead of the chiral basis (i.e., in terms of the couplings $\yctl$ and $\yctr$) as discussed in subsection~\ref{sec:combined analysis:CPV}.

\section{Effective Lagrangian}
\label{sec:EFT}

In order to describe the interactions between the scalar singlet $S$ and the SM fermions, we adopt the EFT approach proposed in refs.~\cite{Franceschini:2016gxv,Banerjee:2018fsx,Castro:2020sba}. After the EW symmetry breaking, the most general effective Lagrangian governing the $tqS$ interaction reads
\begin{align}\label{eq:L:tcS}
    \Lag_{S} \supset -\frac{1}{\sqrt{2}}{S}\big[ 
    \bar{q} (y_S^{qt} + i y_P^{qt} \gamma^5 ) t + \mathrm{h.c.} \big],
\end{align}
where $q=u,\,c$, and the scalar and fermions are all given in the mass eigenbasis. Such interactions arise from the gauge invariant dim-5 operator $(\bar q_L \tilde{H} Y u_R) S$, where $q_L$ denotes the SM left-handed quark doublet, $u_R$ the SM right-handed quark singlet, $\tilde{H}=i\sigma_2 H$ with $H$ being the SM Higgs doublet, and $Y$ the $3 \times 3$ complex matrix in flavour space. Considering the direct searches for the scalar singlet $S$ through the top-quark FCNC decay $t \to c S$ with $S \to \mu^+\mu^- (b \bar b)$ at the LHC, we also include the following flavour-conserving interactions~\cite{Franceschini:2016gxv,Banerjee:2018fsx,Castro:2020sba}:
\begin{align}\label{eq:L:ffS}
    \Lag_{S} \supset -\frac{1}{\sqrt{2}}{S} \big[
    \bar{\mu}(\ymms + i \ymmp\gamma^5 ) \mu 
    + \bar{b}(\ybbs + i \ybbp \gamma^5 ) b\big].
\end{align}
It is noted that the flavour-conserving couplings $y_{S,P}^{\mu\mu}$ and $y_{S,P}^{bb}$ are both real by definition, whereas the top-quark FCNC couplings $y_{S,P}^{qt}$ can be complex in general. Besides the scalar and pseudoscalar basis as defined above, it is also convenient to describe the couplings of the scalar singlet $S$ with the SM fermions in the chiral basis of $y_L^{ij}$ and $y_R^{ij}$, with $y_{R,L}^{ij} \equiv y_S^{ij}\pm iy_P^{ij}$.

The effective Lagrangian in eqs.~\eqref{eq:L:tcS} and \eqref{eq:L:ffS} has been widely adopted in the investigation of direct searches for the top-quark FCNC processes induced by the scalar singlet $S$ at the LHC~\cite{Banerjee:2018fsx,Castro:2020sba,Bahl:2023xkw}. Its UV completion has been studied in refs.~\cite{Banerjee:2018fsx,Batell:2021xsi}. Generally, the scalar $S$ can have other couplings in addition to those introduced in eqs.~\eqref{eq:L:tcS} and \eqref{eq:L:ffS}. For example, considering the $S\tau^+\tau^-$ and $S\gamma\gamma$ couplings, the LHC searches for the top-quark FCNCs with the scalar $S$ through the $S\to \tau^+ \tau^-$ and $S\to\gamma\gamma$ channels have been studied in refs.~\cite{Castro:2020sba} and \cite{Banerjee:2018fsx}, respectively. In addition, the flavour-conserving couplings of $S$ to the quarks can be assessed through the direct searches for the scalar resonances as performed, e.g., in ref.~\cite{CMS:2018pwl}. In this work, we concentrate on the couplings introduced in eqs.~\eqref{eq:L:tcS} and \eqref{eq:L:ffS}. This is motivated by the observation that our main conclusions would not be altered by including other couplings, since they cannot affect the collider processes and the low-energy constraints we are considering here simultaneously. For detailed studies of the flavour structure of the scalar singlet $S$, we refer to refs.~\cite{Froggatt:1978nt,Bauer:2016rxs,Batell:2017kty}.

With the effective interactions specified by eq.~\eqref{eq:L:ffS}, the scalar $S$ can only decay into $b\bar{b}$ and/or $\mu^+\mu^-$. Their decay widths can be written as
\begin{align}\label{eq:h0decay}
    \Gamma(S\to f\bar{f}) =& N_C \frac{m_{S}}{16\pi} \left[\big(y_S^{ff}\big)^2\beta_f^3 + \big(y_P^{ff}\big)^2 \beta_f \right],
\end{align}
with $\beta_f = \sqrt{1-4x_f}$ and $x_f = m_f^2/m_{S}^2$ for $f=b$ or $\mu$. The colour factor $N_C=1$ for $f=\mu$, and $N_C=3$ for $f=b$. In the case of $m_t > m_{S} + m_c$, the top quark can even decay into an on-shell $S$ and a charm quark, with the decay width given by
\begin{align}\label{eq:tdecay}
    \Gamma(t\to c S) &= \frac{3m_t}{32\pi} \Big( \big\lvert \ycts \big\rvert^2+ \big\lvert \yctp \big\rvert^2 \Big) \left(1 - \frac{m_{S}^2}{m_t^2}\right)^2.
\end{align}
Then, the branching ratio can be written as $\mB(t\to c S) \approx \Gamma(t\to c S)/\Gamma(t \to b W)$. For the $t\to b W$ decay, the recent NNLO calculation gives $\Gamma(t\to bW) = 1.33\GeV$~\cite{Chen:2022wit}. When the scalar mass is much higher than the EW scale, the top quark cannot decay into an on-shell $S$. In this case, the effects of the scalar $S$ are adequately described by the four-fermion operators investigated in refs.~\cite{Chala:2018agk,Afik:2021jjh}.

\section{Low-energy constraints}
\label{sec:low-energy constraints}

In this section, we investigate the relevant low-energy processes affected by the scalar singlet $S$, including the $\Bsmumu$ decay, $(g-2)_\mu$ and the neutron EDM.

\subsection[\texorpdfstring{${B_s\to \mu^+\mu^-}$}{Bs → mumu} decay]{$\boldsymbol{\Bsmumu}$ decay}
\label{sec:bsmumu}

\begin{figure}[t]
    \centering
    \subfloat[]{\includegraphics[width=0.25\linewidth]{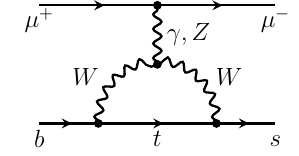}}
    \subfloat[]{\includegraphics[width=0.25\linewidth]{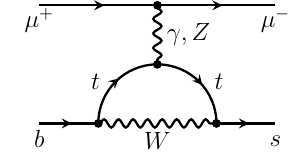}}
    \subfloat[]{\includegraphics[width=0.25\linewidth]{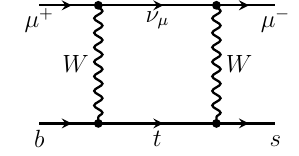}}
    \subfloat[]{\includegraphics[width=0.25\linewidth]{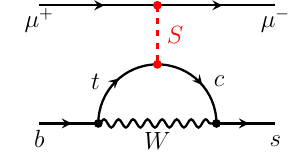}\label{subfig:bsmumu-feynman}}
    \caption{One-loop Feynman diagrams for the $ b\to s\mu^+\mu^- $ transition, including both the dominant SM (the first three) and the scalar $S$ contribution (the last one). }
    \label{fig:bsmumu-feynman}
\end{figure}

The rare decay $\Bsmumu$ is induced by the quark-level $b \to s \mu^+ \mu^-$ transition in the SM, as shown by the first three Feynman diagrams in figure~\ref{fig:bsmumu-feynman}. Due to the helicity, loop and CKM suppression, the decay provides very promising probes of NP effects. With the scalar $S$ contribution taken into account, the low-energy effective Hamiltonian governing the $\Bsmumu$ decay reads~\cite{Buchalla:1995vs,Fleischer:2024fkm}
\begin{align}
    \mathcal{H}_{\text{eff}} \supset 
    -\frac{G_F\alpha_e}{\sqrt{2}\pi} V_{ts}^* V_{tb}  
    \left[ \mC_{10} \mO_{10} + \mC_S \mO_S + \mC_P \mO_P
    \right] +\text{h.c.},
    \label{eq:WET}
\end{align}
where $V_{ij}$ denote the CKM matrix elements. The semi-leptonic four-fermion operators $\mO_i$ are defined, respectively, by
\begin{align}
        \mO_{10} = (\bar{s} \gamma^\mu P_{L} b)(\bar{\mu} \gamma_\mu\gamma_5\mu), \quad 
        \mO_S = m_b(\bar{s}P_{R}b)(\bar{\mu}\mu), \quad 
        \mO_P = m_b(\bar{s}P_{R}b)(\bar{\mu}\gamma_5 \mu),
        \label{eq:operators}
\end{align}
with the chiral projectors given by $P_{L,R}=(1\mp\gamma_5)/2$. In the SM, the Wilson coefficient $\mC_{10}$ arises firstly at the one-loop level, with the contributing Feynman diagrams shown by the first three in figure~\ref{fig:bsmumu-feynman}. Corrections up to the NNLO QCD~\cite{Buchalla:1993bv,Buchalla:1998ba,Misiak:1999yg,Hermann:2013kca} and the NLO EW~\cite{Bobeth:2013tba} have been calculated, which result in $\mC_{10}^\SM=-4.17$ at the scale $\mu=4.8\GeV$~\cite{Bobeth:2013uxa}. In addition, the Wilson coefficients $\mC_{S}$ and $\mC_P$ in the SM are induced by the Higgs-penguin diagrams and are highly suppressed~\cite{Grzadkowski:1983yp,Krawczyk:1989qp,Li:2014fea}. Therefore, they can be safely neglected, i.e., $\mC_S^\SM \approx \mC_P^\SM \approx 0$.

In terms of the effective Lagrangian introduced in section~\ref{sec:EFT}, the $ t c S $ interaction can also affect the $ b\to s \mu^+ \mu^- $ transition through the scalar-penguin diagram shown in figure~\ref{fig:bsmumu-feynman}~\subref{subfig:bsmumu-feynman}. We evaluate this contribution in the unitary gauge with the modified minimal subtraction ($ \overline{\mathrm{MS}} $) scheme. It is found that only the Wilson coefficients $ \mC_{S} $ and $ \mC_P $ are affected and take the form
\begin{align}
\begin{aligned}
    \mC_S^\NP{(\mu_W)} =& \frac{ \ymms \yctr }{8\pi \alpha_e }  \frac{ V_{cs}^* }{ V_{ts}^* } \frac{ m_t }{m_S^2} \cdot F(x_t),  \\
    \mC_P^\NP{(\mu_W)} =& \frac{i \ymmp \yctr }{8\pi \alpha_e } \frac{ V_{cs}^* }{ V_{ts}^* } \frac{ m_t }{m_S^2} \cdot F(x_t) ,
    \label{eq:matching-wcs}
\end{aligned}    
\end{align}
with $ \yctr = \ycts + i\yctp $, $x_t \equiv m_t^2 /m_W^2$ and the loop function given by
\begin{align}
    F(x)=  \frac{1}{2}\ln \frac{\mu_W^2}{m_W^2}+ \frac{ 7-x }{ 4 (1-x) } + \frac{2+2x-x^2  }{2(1-x)^2}\ln x .
\end{align}
It is noted that, compared to $ \mC_{10}^\SM $, both $ \mC_S^\NP $  and $ \mC_P^\NP $ could be largely enhanced by the CKM factor $ V_{cs}^* / V_{ts}^* $.

With the effective Hamiltonian in eq.~\eqref{eq:WET}, the theoretical branching ratio of the $ B_s\to \mu^+\mu^- $ decay can be expressed as~\cite{DeBruyn:2012wk,Buras:2013uqa}
\begin{equation}
    \BR = \frac{G_F^2 \alpha_e^2 }{64\pi^3} \ m_{B_s} \tau_{B_s} f_{B_s}^2 | V_{tb}V_{ts}^* |^2 4 m_\mu^2 \sqrt{1-\frac{4m_\mu^2}{m_{B_s}^2}}\left| \mC_{10}^\SM \right|^2 \left(|P|^2+|S|^2\right), \label{eq:br(bs->mumu)}
\end{equation}
where $ m_{B_s} $, $ \tau_{B_s}=1/\Gamma_s $ and $ f_{B_s} $ denote the mass, lifetime and decay constant of the $ B_s $ meson, respectively. The quantities $ P $ and $ S $ are defined as
\begin{align}\label{eq:PS}
    \begin{aligned}
        P &= 
        1 + \frac{m_{B_s}^2 }{2m_\mu}\left(\frac{m_b}{m_b+m_s}\right) \frac{\mC_P}{\mC_{10}^\SM} 
        \equiv |P|e^{i\varphi_P}, \\
        S &= \sqrt{1-\frac{4m_\mu^2}{m_{B_s}^2}} \frac{m_{B_s}^2 }{2m_\mu}\left(\frac{m_b}{m_b+m_s}\right) \frac{\mC_S}{\mC_{10}^\SM}  \equiv |S|e^{i\varphi_S},
    \end{aligned}
\end{align}
where $ \varphi_P $ and $ \varphi_S $ denote the CP-violating phases of $ P $ and $ S $, respectively.

Taking into account the sizable width difference of the $B_s$ system, the experimentally measured branching ratio of the $\Bsmumu$ decay should be the time-integrated one~\cite{DeBruyn:2012wk,Buras:2013uqa}, which is related to eq.~\eqref{eq:br(bs->mumu)} through
\begin{align}
    \overline{\mB}(\Bsmumu) = \left( \frac{1+\ADG y_s}{1-y_s^2} \right) \BR,
    \label{eq:time-integrated-branching-ratio}
\end{align}
with
\begin{align}
    y_s &= \frac{\Gamma^L_{s}-\Gamma^H_{s}}{\Gamma^L_{s}+\Gamma^H_{s}} = \frac{\Delta\Gamma_s}{2\Gamma_s}.
\end{align}
Here $\Delta\Gamma_s=\Gamma^L_{s}-\Gamma^H_{s}$ is the width difference of the $B_s$ system, and $ \Gamma_s^L $ and $ \Gamma_s^H $ denote the decay widths of the light and heavy $ B_s $ mass eigenstates, respectively. Note that throughout this paper, unless otherwise specified, we use $\BR$ to denote the time-integrated branching ratio defined by eq.~\eqref{eq:time-integrated-branching-ratio}, and confront it with the experimental data given in eq.~\eqref{eq:br_bsmumu:exp}. With the effective interactions in eq.~\eqref{eq:L:tcS}, the scalar singlet $ S $ does not affect the $ B_s$--$\bar{B}_s $ mixing. In this case, the mass-eigenstate rate asymmetry $\ADG $ can be written as
\begin{align}\label{eq:mass-eigenstate-rate-asymmetry}
  \ADG = \frac{|{P}|^2 \cos 2 \varphi_{P} - |{S}|^2 \cos 2 \varphi_{S}}{|{P}|^2  + |{S}|^2},
\end{align}
which is equal to $ +1 $ in the SM but can vary between $-1$ and $+1$ in the presence of NP contributions. $\ADG$ can be obtained via the following relation~\cite{DeBruyn:2012wk,Fleischer:2017yox}:
\begin{align}\label{eq:adg}
    \ADG = \frac{1}{ y_s} \left[ \frac{(1-y_s^2)\tau_{\mu\mu}^s-(1+
 y_s^2)\tau_{B_s}}{2\tau_{B_s}-(1-y_s^2)\tau_{\mu\mu}^s} \right] \ ,
\end{align}
by measuring the effective lifetime $\tau_{\mu\mu}^s$ of the $\Bsmumu$ decay. The current experimental measurements of $\tau_{\mu\mu}^s$ can then be translated into bounds on $\ADG$, yielding
\begin{align}\label{eq:ADG_exp}
  \ADG =
  \begin{cases}
    \;\;\;8.44 \pm 7.11  & \text{LHCb~\cite{LHCb:2021awg}},
    \\
    \;\;\;3.75 \pm 3.68  & \text{CMS~\cite{CMS:2022mgd}},
    \\
          -4.08 \pm 2.52 & \text{ATLAS~\cite{ATLAS:2023trk}},
  \end{cases}
\end{align}
the uncertainties of which are still too large compared to the model-independent range of $-1 \leq \ADG \leq +1$.

We can also consider the fully time-dependent and flavour-tagged rate of the $ \Bsmumu $ decay with specific helicity configurations of the two muon final states. Then, the helicity-dependent rate asymmetry can be defined as~\cite{DeBruyn:2012wk,Buras:2013uqa}
\begin{equation}\label{eq:ACP:with helicity}
    \frac{\Gamma(B^0_s(t) \to \mu_\lambda^+ \mu_\lambda^-) - \Gamma(\bar B^0_s(t) \to \mu_\lambda^+ \mu_\lambda^-)}{\Gamma(B^0_s(t) \to \mu_\lambda^+ \mu_\lambda^-) + \Gamma(\bar B^0_s(t) \to \mu_\lambda^+ \mu_\lambda^-)} = \frac{\Cmumulambda \cos (\Delta M_s t) + \Smumu \sin (\Delta M_s t)}{\cosh (y_s t/\tau_{B_s}) + \ADG \sinh (y_s t/\tau_{B_s})} \ ,
\end{equation}
where $\Delta M_s$ denotes the mass difference between the heavy and light $B_s$ mass eigenstates. The helicity eigenvalues of the muon pair are given by $ \eta_L = +1 $ and $ \eta_R = -1 $ for $\lambda = L$ and $\lambda = R$, respectively. Making using of eq.~\eqref{eq:PS}, the two CP observables $\Cmumulambda$ and $\Smumu$ can be written as
\begin{align}
    \Cmumulambda &= -\eta_\lambda \frac{2 |{P S}| \cos(\varphi_P - \varphi_S)}{|{P}|^2 + |{S}|^2} \equiv -\eta_\lambda \Cmumu \ , \label{eq:Cmumu}  \\ 
    \Smumu &= \frac{|{P}|^2 \sin 2 \varphi_P - |{S}|^2 \sin 2 \varphi_S }{|{P}|^2 + |{S}|^2} \ ,
    \label{eq:Smumu}
\end{align}
both of which are equal to zero in the SM, but lie between $-1$ and $+1$ in the presence of NP contributions. Summing eq.~\eqref{eq:ACP:with helicity} over the muon helicities, we obtain the time-dependent rate asymmetry
\begin{equation}\label{eq:ACP:sum over helicity}
    \frac{\Gamma(B^0_s(t) \to \mu^+ \mu^-) - \Gamma(\bar B^0_s(t) \to \mu^+ \mu^-)}{\Gamma(B^0_s(t) \to \mu^+ \mu^-) + \Gamma(\bar B^0_s(t) \to \mu^+ \mu^-)} = \frac{\Smumu \sin (\Delta M_s t)}{\cosh (y_s t/\tau_{B_s}) + \ADG \sinh (y_s t/\tau_{B_s})}.
\end{equation}
Comparing the two CP-violating rate asymmetries defined by eqs.~\eqref{eq:ACP:with helicity} and \eqref{eq:ACP:sum over helicity}, we can see that the observables $\ADG$ and $\Smumu$ can be extracted without measuring the muon helicity, while $\Cmumulambda$ is difficult to measure, since it requires the information of the muon helicity. It should be noted that the three observables $ \ADG $, $ \Smumu $ and $ \Cmumu $ are not independent of each other, but satisfy the normalization relation
\begin{align}
    (\ADG)^2 + (\Smumu)^2 + (\Cmumulambda)^2 = 1 \ .
    \label{eq:sphere}
\end{align}
Therefore, we choose $\ADG$ and $\Smumu$ as the two independent CP observables in the following analysis. They are also theoretically cleaner than the branching ratio, since their dependence on the decay constant $f_{B_s}$ is cancelled.

\begin{table}
  \tabcolsep 0.15in
  \renewcommand*{\arraystretch}{1.2}
  \centering
  \begin{tabular}{c c c c}
    \toprule
    Input & Value & Unit & Ref. \\ \midrule
    $ \tau_{B_s} $ & $ 1.520 \pm 0.005  $ & $10^{-12}$\,s & ~\cite{ParticleDataGroup:2024cfk}\\
    $ \Delta\Gamma_s/\Gamma_s $ & $ 0.127\pm 0.007 $ &  & ~\cite{ParticleDataGroup:2024cfk}\\
    \midrule
    $m_t^{\text{pole}}$ & $172.57 \pm 0.29$ & $\GeV$ & ~\cite{ParticleDataGroup:2024cfk} \\
    $ m_b(m_b) $ & $ 4.183 \pm 0.007 $ & $ \GeV $ & ~\cite{ParticleDataGroup:2024cfk} \\
    $ m_s(m_s) $ & $ 93.5 \pm 0.8 $ & $ {\rm MeV} $ & ~\cite{ParticleDataGroup:2024cfk}\\
    \midrule
    $ f_{B_s} |_{N_f=2+1+1}$ & $230.3 \pm 1.3 $ & ${\rm MeV}$ & ~\cite{FlavourLatticeAveragingGroupFLAG:2021npn}\\         $f^{K \to \pi}_+(0)$  & $0.9698 \pm 0.0017$ &  & ~\cite{FlavourLatticeAveragingGroupFLAG:2021npn}\\                  \midrule
    $|V_{us}|f^{K \to \pi}_+$(0) & $0.21635 \pm 0.00038 $ & & ~\cite{Charles:2004jd, Moulson:2017ive}
    \\
    $|V_{ub}|$ & $ 3.86 \pm 0.07 \pm 0.12  $ & $10^{-3}$ &  ~\cite{Charles:2004jd}
    \\
    $|V_{cb}|$ & $ 41.22 \pm 0.24 \pm 0.37 $& $10^{-3}$ &  ~\cite{Charles:2004jd}
    \\
    $\gamma$& $66.4^{+2.8}_{-3.0}$ & ${}^\circ$ & ~\cite{HFLAV:2022esi} 
    \\\bottomrule
  \end{tabular}
  \caption{Main input parameters used in our numerical analysis.}
  \label{tab:input}
\end{table}

In table~\ref{tab:input}, we list the main input parameters used in our numerical analysis. With these input values, our SM prediction of $\BR$ has been already given in eq.~(\ref{eq:br_bsmumu:SM}).\footnote{For the input values of $|V_{ub}|$ and $|V_{cb}|$, we have adopted the CKMfitter average of the inclusive and exclusive determinations as given in table~\ref{tab:input}. When taking the inclusive (exclusive) values as input, the SM prediction becomes $5\%$ higher ($10\%$ lower) than our prediction in eq.~(\ref{eq:br_bsmumu:SM}), which has only a minor impact on our subsequent numerical analysis of the NP effects. For more detailed discussions about the choices of $|V_{ub}|$ and $|V_{cb}|$, we refer to refs.~\cite{Bobeth:2021cxm,DeBruyn:2022zhw,Fleischer:2024fkm}.} Taking further into account the scalar $S$ contribution (cf. eq.~\eqref{eq:matching-wcs}), the branching ratio of $\Bsmumu$ can be numerically written as\footnote{In order to present a compact numerical form, the correction from $\ADG$ is calculated by taking $\varphi_S=\varphi_P=0$. In our numerical analysis, however, the full analytical expressions are always used.}
\begin{align}\label{eq:Bsmumu:simi-num}
    \BR \times 10^9 
    =\ 3.51 &- 3.34\ {\rm Im}\left(\frac{\ymmp \yctr}{0.01}\right)
        \left(\frac{100\GeV}{m_S}\right)^2 \\
        & + \biggl ( 0.79\left|\frac{\ymmp \yctr}{0.01}\right|^2 
        + 0.70 \left|\frac{\ymms \yctr}{0.01}\right|^2 \biggr )
        \left(\frac{100\GeV}{m_{S}} \right)^{4}. \nonumber
\end{align}
It can be seen that the effects of the $tcS$ couplings are largely enhanced compared to the SM contributions. This can be understood from the fact that the corresponding Wilson coefficients $\mC_S^\NP$ and $\mC_P^\NP$ contain a large CKM factor (cf. eq.~\eqref{eq:matching-wcs}) and their contributions to the branching ratio do not suffer from the helicity suppression (cf. eqs.~\eqref{eq:br(bs->mumu)} and \eqref{eq:PS}). In addition, the difference between the coefficients of the $|\ymmp\yctr|^2$ and $|\ymms\yctr|^2$ terms arise from the effects of $\ADG$.

\begin{figure}[t]
  \centering
  \includegraphics[width=0.49\linewidth]{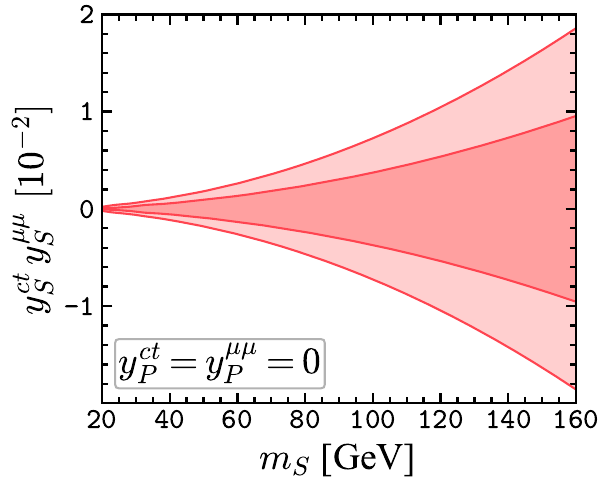}\;
  \includegraphics[width=0.49\linewidth]{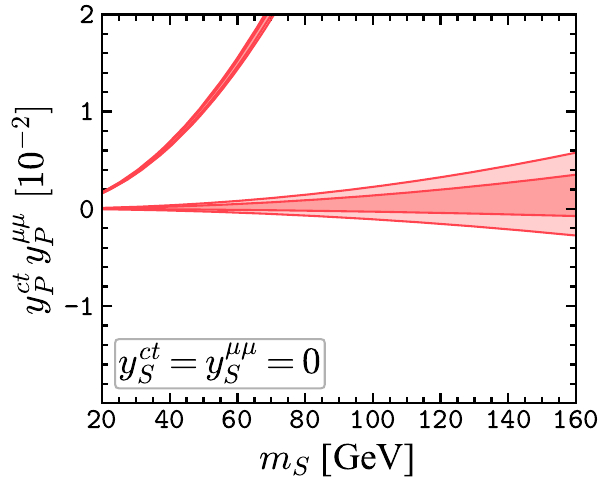}
  \caption{Constraints on $\ycts\ymms$ (left) and $\yctp\ymmp $ (right) as a function of $m_S$ from the $\Bsmumu$ decay, in which $\yctp=\ymmp=0$ and $\ycts=\ymms=0$ are assumed, respectively. The allowed regions at the $1\sigma$ and the $2\sigma$ confidence level~(CL) are shown by the dark and the light red, respectively.}
  \label{fig:prediction}
\end{figure}

To illustrate the constraint from the measured branching ratio $\BR$, we consider two simplified cases of $\yctp=\ymmp =0$ and $\ycts=\ymms=0$, and show in figure~\ref{fig:prediction} the allowed parameter regions in the $(m_S,\, \ycts\ymms)$ and $(m_S, \, \yctp\ymmp)$ planes, respectively. In the former case, there exists only one allowed parameter region, since the NP contribution to $\mC_S$ has no linear interference with the dominant SM contribution $\mC_{10}^\SM$ and appears firstly at the quadratic order, as can be seen from eqs.~\eqref{eq:br(bs->mumu)} and \eqref{eq:Bsmumu:simi-num}. In the latter case, on the other hand, there exist two allowed parameter regions: in the region with larger values of $\yctp\ymmp$, the sign of the quantity $P$ (cf. eq.~(\ref{eq:PS})) is flipped by the large NP contribution to $\mC_P$; in the other region, the allowed values of $\yctp\ymmp$ are much smaller and give a constructive interference with the dominant SM contribution $\mC_{10}^\SM$.

\subsection[Muon \texorpdfstring{${g-2}$}{g-2}]{Muon $\boldsymbol{g-2}$}
\label{sec:g-2}

The muon anomalous magnetic moment $a_\mu=(g-2)_\mu/2$ can provide powerful probes of NP effects~\cite{Jegerlehner:2009ry,Aoyama:2020ynm}. Taking into account the latest measurement by the Fermilab Muon $g-2$ Experiment~\cite{Muong-2:2023cdq}, the combined experimental average $a_\mu^\text{Exp}$ shows a $ 5.1\sigma $ discrepancy with the community-approved SM prediction $a_\mu^{\SM,\,\text{WP}}$ from the Muon $g-2$ Theory Initiative~\cite{Aoyama:2020ynm},
\begin{align}
    \Delta a_\mu^{\text{WP}}=a_\mu^\text{Exp}-a_\mu^{\SM,\,\text{WP}}=\left(249\pm 48\right)\times 10^{-11},
    \label{eq:muong-2}
\end{align} 
which implies the presence of potential NP effects. For the SM prediction from the Muon $g-2$ Theory Initiative~\cite{Aoyama:2020ynm}, the hadronic vacuum polarization (HVP) contribution is evaluated by using the $e^+ e^- \to \text{hadrons}$ cross sections measured by multiple experiments. However, a recent measurement of the $e^+ e^- \to \pi^+ \pi^-$ cross section from the CMD-3 experiment is found to be significantly higher than all the previous measurements~\cite{CMD-3:2023alj,CMD-3:2023rfe}, which results in good consistency between $a_\mu^{\text{Exp}}$ and $a_\mu^\SM$. On the other hand, the lattice calculation of the HVP contribution by the BMW collaboration also disagrees significantly with the $e^+ e^-$ data~\cite{Borsanyi:2020mff}, but goes in the same direction as the CMD-3 result. Recently, a new lattice calculation of the HVP contribution is performed in ref.~\cite{Boccaletti:2024guq}, leading to
\begin{align}\label{eq:muong-2:2}
    \Delta a_\mu^{\text{Lattice}} = a_\mu^\text{Exp}-a_\mu^{\SM,\,\text{Lattice}}=\left(40\pm 44\right)\times 10^{-11},
\end{align} 
which reduces the discrepancy to be only of $0.9$ standard deviation. Although continued efforts are still needed to clarify the current theoretical situation~\cite{Colangelo:2022jxc}, these recent evaluations of the HVP contribution also motivate us to consider the possibility of a good consistency between $a_\mu^{\text{Exp}}$ and $a_\mu^\SM$. Therefore, both $\Delta a_\mu^{\text{WP}}$ in eq.~(\ref{eq:muong-2}) and $\Delta a_\mu^{\text{Lattice}}$ in eq.~(\ref{eq:muong-2:2}) will be considered in our numerical analysis.

\begin{figure}[t]
    \centering
    \includegraphics[width=\linewidth]{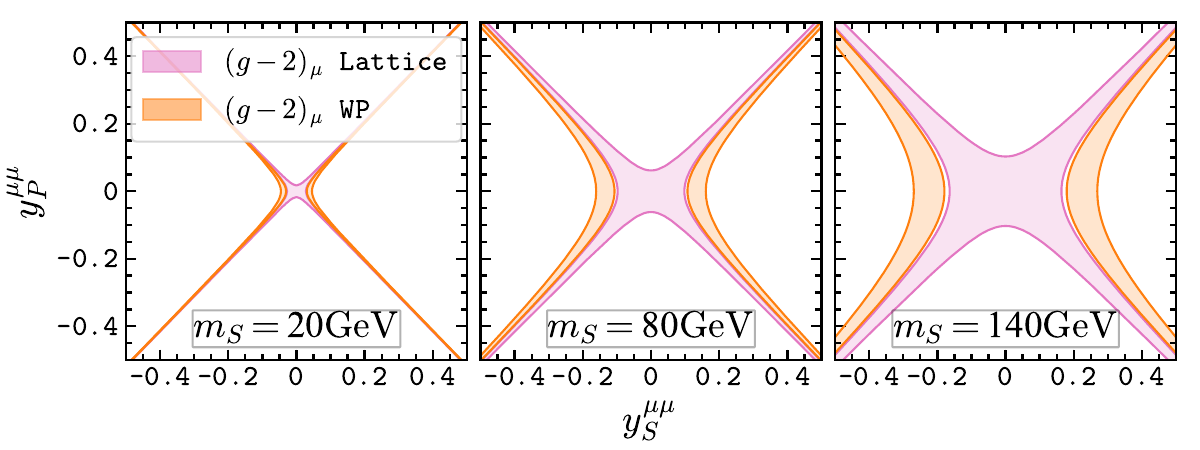}
    \caption{Constraints on $(\ymms,\, \ymmp)$ from the $(g-2)_\mu$ at the $2\sigma$ level, for three benchmark masses $m_S=20$, $80$ and $140\GeV$. The orange and pink regions are allowed by considering eqs.~(\ref{eq:muong-2}) and (\ref{eq:muong-2:2}), respectively.}
    \label{fig:muong-2}
\end{figure}

With the effective interactions given in eq.~\eqref{eq:L:ffS}, the scalar $S$ can shift $a_\mu$ by~\cite{Queiroz:2014zfa}
\begin{align}
 \Delta a_\mu^\NP &= \frac{1}{8\pi^2}\frac{m_\mu^2}{m_{S}^2} \left[\left(\ymms\right)^2 \left( \ln\frac{m_{S}}{m_\mu} - \frac{7}{12} \right)  + \left(\ymmp \right)^2 \left( -\ln\frac{m_{S}}{m_\mu} + \frac{11}{12}\right) \right] \ ,
\end{align}
in the regime $m_S\gg m_\mu$. In figure~\ref{fig:muong-2}, the parameter regions in the $(\ymms,\, \ymmp)$ plane allowed to explain the $(g-2)_\mu$ are shown for three benchmark masses $m_S=20$, $80$ and $140\GeV$. Considering $\Delta a_\mu^{\text{WP}}$ in eq.~(\ref{eq:muong-2}), we can see that the coupling $\ymms$ should be nonzero, since the contribution related to $\ymmp$ is always negative and thus increases the discrepancy of $(g-2)_\mu$. Confronted with $\Delta a_\mu^{\text{Lattice}}$ in eq.~(\ref{eq:muong-2:2}), on the other hand, both $\ymms$ and $\ymmp$ can be zero within the $2\sigma$ allowed regions.

\subsection{Neutron EDM}

The scalar singlet $S$ with complex $tcS$ couplings given in eq.~\eqref{eq:L:tcS} could also affect the hadronic EDM~\cite{Blankenburg:2012ex,Harnik:2012pb,Gorbahn:2014sha}. Generally, the low-energy effective Hamiltonian for the hadronic EDM takes the form~\cite{Gorbahn:2014sha}
\begin{align}
    \begin{aligned}
        \mathcal H_{\text{eff}} \supset & +d_q(\mu) \cdot \frac{i}{2} \bar{q} \sigma^{\mu \nu} \gamma_5 q F_{\mu \nu}+\tilde{d}_q(\mu) \cdot \frac{i}{2} g_s(\mu) \bar{q} \sigma^{\mu \nu} T^a \gamma_5 q G_{\mu \nu}^a \\
        & +w(\mu) \cdot \frac{1}{3} f^{a b c} G_{\mu \sigma}^a G_\nu^{b, \sigma} \widetilde{G}^{c, \mu \nu},
        \label{eq:Heff:EDM}
    \end{aligned}
\end{align}
where the index $q$ runs over all quarks lighter than the top-quark mass $m_t$, and $g_s(\mu)=\sqrt{4\pi\alpha_s(\mu)}$ is the QCD gauge coupling. $F_{\mu\nu}$ and $G_{\mu\nu}^a$ denote the field-strength tensors of QED and QCD respectively, while $\widetilde{G}^{a, \mu \nu}=\frac{1}{2} \epsilon^{\mu \nu \alpha \beta} G_{\alpha \beta}^a$ is the dual field-strength tensor of QCD, with the fully anti-symmetric Levi-Civita tensor $\epsilon^{\mu\nu\lambda\rho}$ defined with the convention $\epsilon^{0 1 2 3}=+1$. $T^a$ and $f^{abc}$ are the generators and structure constants of the gauge group $SU(3)_C$, respectively.

\begin{figure}[t]
  \centering
  \includegraphics[width=0.9\linewidth]{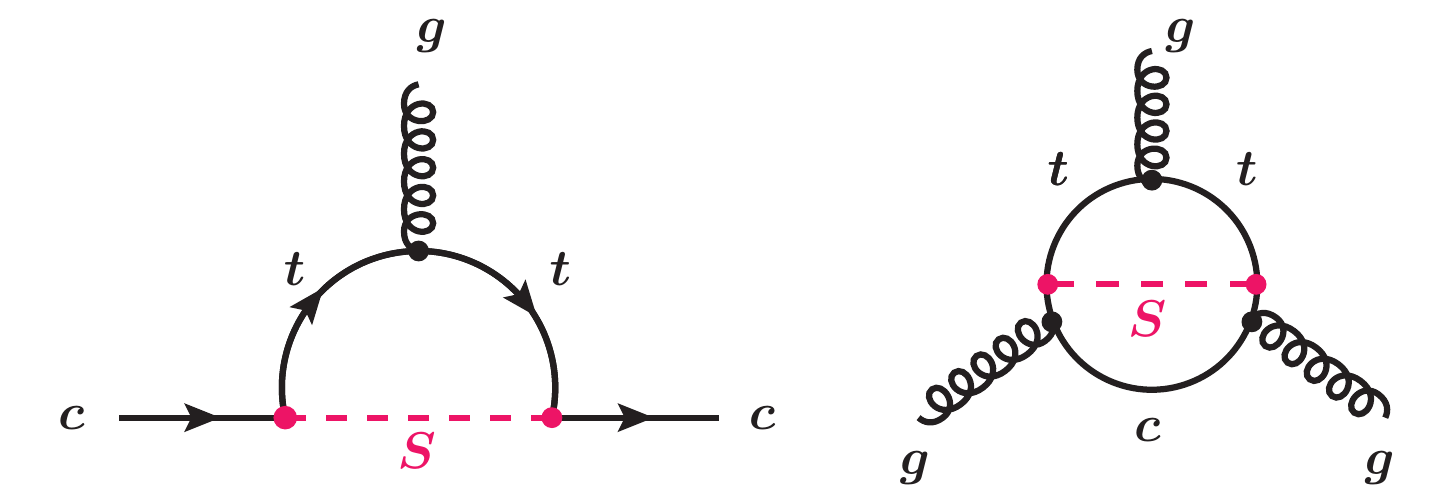}
  \caption{Left: One-loop Feynman diagram contributing to the CEDM of the charm quark. Right: Example of a two-loop Feynman diagram inducing the Weinberg operator. Both are due to the scalar singlet $S$ with complex $tcS$ couplings in eq.~\eqref{eq:L:tcS}.}
  \label{fig:CEDM}
\end{figure}

The leading contribution to the charm-quark chromoelectric dipole moment (CEDM) $\tilde{d}_c$ at the matching scale $\mu_{t}=\mO(m_t)$ is obtained from the one-loop Feynman diagram shown on the left-hand side of figure~\ref{fig:CEDM}. After integrating out the top quark and the scalar boson, it is straightforward to obtain~\cite{Boyd:1990bx,Jung:2013hka,Gorbahn:2014sha}
\begin{align}\label{eq:WC:CEDM:mt}
    \tilde{d}_c(\mu_t)=\frac{1}{32 \pi^2} \frac{m_t}{m_S^2} f_1(x_{tS}) \operatorname{Re}(\ycts \yctp^*),
\end{align}
with $x_{tS}=m_t^2/m_S^2$ and the loop function
\begin{align}\label{eq:f1_function_for_CEDM}
    f_1(x)=\frac{x-3}{(x-1)^2}+\frac{2}{(x-1)^3}\ln{x}.
\end{align}
One can use the relation $\operatorname{Re}(\ycts \yctp^*)= \operatorname{Im}(\yctr \yctl^*)/2$ to express eq.~(\ref{eq:WC:CEDM:mt}) in terms of $\yctr$ and $\yctl$. The scalar singlet $S$ can also contribute to the charm-quark EDM $d_c$ through the same one-loop Feynman diagram in figure~\ref{fig:CEDM} but with the gluon replaced by a photon. As a result, one obtains $d_c(\mu_t)= Q_c e \tilde d_c(\mu_t)$, where $Q_c=+2/3$ is the charm-quark electric charge. The charm-quark EDM affects the neutron EDM through the flavour-mixing contributions into the down-quark EDM~\cite{Cordero-Cid:2007cmf,Grozin:2009jq} and mixing into the charm-quark CEDM under the QED Renormalization Group Evolution~(RGE)~\cite{Gisbert:2019ftm,Ema:2022pmo}. However, the contribution from $d_c$ is found to be smaller than from $\tilde d_c$ by at least one order of magnitude~\cite{Gisbert:2019ftm}. Therefore, the effect from the charm-quark EDM $d_c$ can be safely neglected.

The matching condition $w(\mu_t)$ of the Weinberg operator is obtained by computing the two-loop Feynman diagrams in the full theory~\cite{Weinberg:1989dx,Dicus:1989va}, an example of which is displayed on the right-hand side of figure~\ref{fig:CEDM}. Resorting to the results presented in refs.~\cite{Jung:2013hka,Gorbahn:2014sha} and setting the charm-quark mass to zero, we can write its explicit expression as
\begin{align}\label{eq:WC:Weinberg:mt}
    w(\mu_t)=\frac{g_s^3(\mu_t)}{(32 \pi^2)^2} \frac{1}{m_S^2} f_2(x_{tS}) \operatorname{Re}(\ycts \yctp^*),
\end{align}
with the loop function given by
\begin{align}\label{eq:f2_function_for_Weinberg_operator}
    f_2(x)=-\frac{x^2 -5x -2}{3(x-1)^3}-\frac{2x}{(x-1)^4}\ln{x}.
\end{align}
We can also express eq.~(\ref{eq:WC:Weinberg:mt}) in terms of $\yctr$ and $\yctl$ by making use of the relation $\operatorname{Re}(\ycts \yctp^*)= \operatorname{Im}(\yctr \yctl^*)/2$.

The Wilson coefficients in the effective Hamiltonian of eq.~(\ref{eq:Heff:EDM}) at the hadronic scale $\mu_{H}=1\GeV$ can be obtained by performing the necessary RGE~\cite{Degrassi:2005zd}. In the leading-logarithmic approximation, their numerical expressions are given by~\cite{Gorbahn:2014sha}
\begin{align}
    \begin{aligned}
      \frac{d_d(\mu_H)}{e}&=2.3 \times 10^{-8} \cdot \tilde{d}_c(\mu_t)+1.0 \times 10^{-4} \GeV \cdot w(\mu_t), \\
      \frac{d_u(\mu_H)}{e}&=-2.1 \times 10^{-8} \cdot \tilde{d}_c(\mu_t)-9.1 \times 10^{-5} \GeV \cdot  w(\mu_t), \\[0.2em]
      \tilde{d}_d(\mu_H)&=1.8 \times 10^{-6} \cdot \tilde{d}_c(\mu_t)+7.0 \times 10^{-4} \GeV \cdot w(\mu_t), \\[0.5em]
      \tilde{d}_u(\mu_H)&=8.2 \times 10^{-7} \cdot \tilde{d}_c(\mu_t)+3.1 \times 10^{-4} \GeV \cdot w(\mu_t), \\[0.5em]
      w(\mu_H)&=1.7 \times 10^{-2} \GeV^{-1} \cdot \tilde{d}_c(\mu_t)+0.41 \cdot w(\mu_t).
      \label{eq:WC:EDM:1GeV}
    \end{aligned}
\end{align}
It is noted that, under renormalization, the Weinberg operator mixes into the quark EDMs and CEDMs, but the opposite is not true. The appearance of the charm-quark CEDM $\tilde d_c$ in the above equations is due to the charm-quark threshold correction to $w(\mu)$ at the scale $\mu=m_c$~\cite{Sala:2013osa,Boyd:1990bx,Braaten:1990gq,Chang:1990jv} and the subsequent RGE to the scale $\mu_H=1\GeV$. It can be seen that the RGE always tends to suppress these Wilson coefficients when running from a high down to a low scale.

In terms of the Wilson coefficients evaluated at the scale $\mu_H$, the neutron EDM takes the following form~\cite{Pospelov:2005pr}:
\begin{align}
        \frac{d_n}{e}= & (1.0 \pm 0.5) \Biggl\{ 1.4\left[\frac{d_d(\mu_H)}{e}-0.25 \frac{d_u(\mu_H)}{e}\right ] + 1.1 \left[ \tilde{d}_d(\mu_H)+0.5 \tilde{d}_u(\mu_H) \right] \Biggr\} \nonumber \\
        & +(22 \pm 10) \times 10^{-3} \GeV \cdot w (\mu_H). \label{eq:EDM:num:mH}
\end{align}
Plugging the results in eq.~(\ref{eq:WC:EDM:1GeV}) into the above equation, we can express the neutron EDM in terms of the Wilson coefficients evaluated at the scale $\mu_t$ as
\begin{align}
    \left\lvert\frac{d_n}{e}\right\rvert=\left\lvert (3.7\pm 1.7) \times 10^{-4} \cdot \tilde{d}_c(\mu_t)+ (9.3\pm 4.1) \times 10^{-3} 
    \GeV \cdot w (\mu_t)\right\rvert.
    \label{eq:EDM:num:mt}
\end{align}
As an illustration, taking $m_S=100\GeV$ and together with eqs.~\eqref{eq:WC:CEDM:mt} and \eqref{eq:WC:Weinberg:mt}, this translates into 
\begin{equation}
\left\lvert\frac{d_n}{e}\right\rvert = (1.11\pm 0.51)\times 10^{-22}\,\text{cm} \cdot |\operatorname{Re} (\ycts \yctp^*)| .
\end{equation}

\begin{figure}
  \centering
  \includegraphics[width=0.5\linewidth]{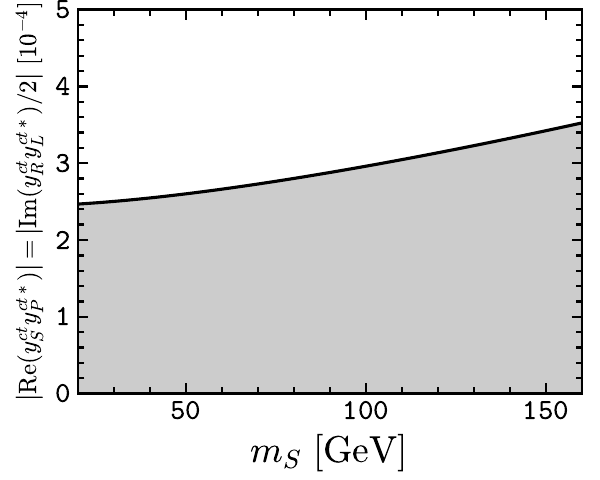}
  \caption{$90\%$ CL upper bound on $|\operatorname{Re}(\ycts \yctp^*)|$ or $|\operatorname{Im}(\yctr \yctl^*)/2|$ as a function of $ m_S $ from the neutron EDM.}
  \label{fig:EDM}
\end{figure}

The current $90\%$ CL upper limit on the neutron EDM reads~\cite{Abel:2020pzs}
\begin{align}
    \left|\frac{d_n}{e}\right|< 1.8 \cdot 10^{-26} \,\text{cm}.
    \label{eq:limitation_for_EDM}
\end{align}
We can use this limit to derive constraint on the $tcS$ couplings, which is shown in figure~\ref{fig:EDM}. Here, to be conservative, we have set the numerical coefficients $(1.0 \pm 0.5)$ and $(22 \pm 10)$ in eq.~\eqref{eq:EDM:num:mH} to $0.5$ and $12$, respectively. It can be seen that the upper bound on $|\operatorname{Re}(\ycts \yctp^*)|$ or $|\operatorname{Im}(\yctr \yctl^*)/2|$ is of the order of $10^{-4}$ and depends weakly on the scalar mass $m_S$.

\section{Phenomenology at the LHC}
\label{sec:collider}

\begin{figure}[t]
  \centering
  \subfloat[$t \bar t$ production with one top decaying into $ qS $.]
  {\includegraphics[width=0.48\textwidth]{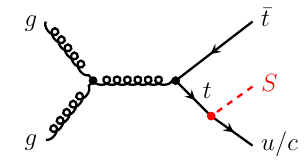}}
  \hspace{0.2cm}
  \subfloat[single-top production associated with an $S$.]
  {\includegraphics[width=0.48\textwidth]{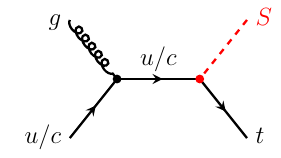}}
  \caption{Representative Feynman diagrams for the production channels of the scalar $S$ at the LHC, with its FCNC couplings to the top quark specified by eq.~\eqref{eq:L:tcS}.}
  \label{fig:diagram:LHC}
\end{figure}

In this section, we investigate the collider signatures of the scalar singlet $S$ introduced in section~\ref{sec:EFT}. According to its FCNC couplings to the top quark (cf. eq.~\eqref{eq:L:tcS}), the scalar $S$ can be produced via the following two channels at the LHC: one is the top-pair production with one of the top quarks decaying into $ qS $ ($q=u,\, c$), and the other is the single-top production associated with a scalar $S$, i.e., $ q g\to t S $. In figure~\ref{fig:diagram:LHC}, we show the corresponding Feynman diagrams, where the charge-conjugated processes are not shown but will be included in our numerical analysis. In subsections~\ref{sec:LHC:mumu} and \ref{sec:LHC:bb}, we focus on the scenarios in which the scalar $S$ decays to $\mu^+ \mu^-$ and $b \bar b$ pair, respectively. The scenario in which both of the two decay channels are involved will be considered in subsection~\ref{sec:LHC:combined}. In addition, since the processes we are considering are not sensitive to the difference between the couplings $\ycts$ and $\yctp$, we can take, for simplicity, $\yctp =0$ in the following analysis.

In our numerical analysis, the signal and the SM background events are simulated with \texttt{MadGraph5}~\cite{Alwall:2014hca}. Dedicated Universal FeynRules Output (UFO)~\cite{Degrande:2011ua} model files are produced with \texttt{FeynRules}~\cite{Alloul:2013bka}. Then, \texttt{Pythia8}~\cite{Bierlich:2022pfr} is used to simulate the parton shower and hadronization, and \texttt{Delphes}~\cite{deFavereau:2013fsa} is used to simulate the detector effects with the default CMS card, in which the anti-$ k_t $ algorithm~\cite{Cacciari:2008gp} is chosen for clustering the jets with radius parameter $ \Delta R = 0.4 $. Finally, the detector-level samples are analyzed with the \texttt{MadAnalysis5}~\cite{Conte:2012fm}. In addition, we use the NNPDF2.3~\cite{Ball:2012cx} leading-order parton distribution functions during our calculation.

\begin{table}[t]
  \begin{center}
  \resizebox{\textwidth}{!}{
    \renewcommand*{\arraystretch}{1.2}
    \begin{tabular}{c|ccc|ccc|cc}
      \bottomrule
      &\multicolumn{3}{c|}{Top pair} & \multicolumn{3}{c|}{Single top} &  \multicolumn{2}{c}{Others}\\
      \midrule
      & $tt $ & $ttbb$ & $ttV$ & $tj$ & $tW$ & $tb$ & $tZ$ & $VV$ \\
      \midrule
      $K$-factor &  1.6  & 1.25   &  1.3 & 1.2  & 1.3  & 1.5 & 1.6 & 1.7 \\
      \midrule
      Refs. & ~\cite{Czakon:2013goa,Czakon:2011xx} & ~\cite{Bredenstein:2010rs} & ~\cite{Campbell:2012dh} & ~\cite{Kidonakis:2018ncr} & ~\cite{Kidonakis:2018ncr} & ~\cite{Kidonakis:2018ncr} & ~\cite{Campbell:2013yla} & ~\cite{Campbell:2011bn} \\
      \toprule
    \end{tabular}}
    \caption{The $ K $-factors for the different background channels.}
    \label{tab:K-factors}
  \end{center}
\end{table}

The generated signal events for the top-pair production channel are normalized by using the NNLO+NNLL theoretical prediction $ \sigma(pp\to t\bar{t})=984.5\pb $ at $\sqrt{s}=14\TeV$~\cite{Czakon:2013goa} and the experimental measurement $ \mB (t\to b \ell \nu_\ell) =0.225 $ with $\ell=e,\,\mu$~\cite{ParticleDataGroup:2024cfk}, while the cross section for $ p p\to t S $ channel is calculated at the leading order in QCD. In order to include the higher-order QCD corrections for the background channels, the simulated samples generated at the leading order are scaled by the $K$-factors, which are defined as the ratio of the higher- and the leading-order cross-sections calculated. For convenience, their values are listed in table~\ref{tab:K-factors} for the different background channels.

In order to obtain the upper limits on the signal cross-sections, we adopt the statistical significance defined in ref.~\cite{Cowan:2010js}
\begin{align}
    Z=\sqrt{2\left[(S + B) \ln\left(1+ \frac{S}{B}\right) - S\right]},
    \label{eq:z-test}
\end{align} 
where $ S $ and $ B $ denote the total event numbers in a specific phase-space region of the signal and background channels, respectively. They can be evaluated by $ \mL \times \sigma \times \mB \times \varepsilon $, in which $ \mL $ is the integrated luminosity, and $ \sigma $ and $ \mB $ denote the cross section and the decay branching fraction of a specific channel, respectively. $ \varepsilon $ represents the cut efficiency for a specific signal region, which will be studied in detail for the decay channels $S\to \mu^+\mu^-$ and $S\to b\bar{b}$ in subsection~\ref{sec:LHC:mumu} and \ref{sec:LHC:bb}, respectively.

\subsection[Three-lepton final states through \texorpdfstring{${S\to \mu^+\mu^-}$}{H0 -> mu mu} channel]{Three-lepton final states through $\boldsymbol{S\to \mu^+\mu^-}$ channel}
\label{sec:LHC:mumu}

In this subsection, we focus on the scenario in which the scalar $S$ decays into a pair of muons. For the two production channels shown in figure~\ref{fig:diagram:LHC}, we consider the following decay chains:
\begin{align*}
    p p & \to t(\to b\ell \nu_\ell) t(\to q S, S\to \mu^+\mu^-) && \to 3\ell + 1b + 1j + \slashed{E}_T, \\
    p p & \to t(\to b\ell \nu_\ell) S(\to \mu^+\mu^-)           && \to 3\ell + 1b + \slashed{E}_T,
\end{align*}
where $b$ and $j$ denote the $b$-tagged and the light jet, while $\slashed{E}_T$ is the missing transverse energy. The signal contains exactly three isolated leptons (two of them are muons with opposite charges), at least one jet (one is the $b$-tagged jet), and missing transverse energy.

The same signature can also arise from multiple SM backgrounds. Here we consider only the most significant ones arising from $ t\bar{t} $, $ t\bar{t}V $, $ tZ $, $ VV $ and $ tW $, with $ V = W,Z $, together with their detailed decay chains given by
\begin{align*}
    t\bar{t}:  && pp &\to t (\to b \ell \nu_\ell ) \bar{t}(\to \bar{b} \ell \bar{\nu}_l ) &&\to 2\ell + 2b + \slashed{E}_T, \\
    t\bar{t}V: && pp &\to t(\to b \ell \nu_\ell ) \bar{t}(\to \bar{b} \ell \bar{\nu}_l ) V(\to \ell+\ell/\nu_\ell) &&\to \text{3-4}\ell + 2b + \slashed{E}_T, \\
    tZ:        && pp &\to t(\to b \ell \nu_\ell ) Z(\to \ell \ell) j &&\to 3\ell + 1b + 1j + \slashed{E}_T, \\
    VV:        && pp &\to V V(\to \ell+\ell/\nu_\ell) && \to \text{2-4}\ell + \slashed{E}_T, \\
    tW:        && pp &\to t(\to b \ell \nu_\ell ) W(\to \ell \nu_\ell ) &&\to 2\ell + 1b + \slashed{E}_T,
\end{align*}
where the CP-conjugated processes are self-evident.

The simulated signal and background samples are generated following the steps described above. Furthermore, the following fiducial phase-space selections at the parton level are imposed:
\begin{align}
     p_T^{j} &> 20 \GeV, & p_T^\ell &> 10 \GeV, & |\eta^{j}| &< 5, & |\eta^\ell| &< 2.5, & \Delta R(m, n) &> 0.4~(m, n = \ell, b, j),
\end{align}
where $ p_T $ and $ \eta $ denote the transverse momentum and the pseudo-rapidity, respectively. $ \Delta R(m,n) = \sqrt{(\Delta\eta)^2+(\Delta\phi)^2} $ denotes the separation between the particles $m$ and $n$ in the $\eta-\phi$ plane. In addition, since the values of $ tcS $ couplings do not affect the distributions of final states, we take $\ycts=0.001$ for the signal sample simulation.

At the detector level, the jets are required to have $ p_T > 25 \GeV$ and $ |\eta|< 2.5 $, while the leptons should have $ p_T > 15 \GeV$ and $ |\eta| < 2.5 $ and at least one lepton is required to have $ p_T > 25 \GeV$ to satisfy the trigger threshold. With these prerequisites, we select signals by further requiring the events to contain exactly three isolated leptons (two of them are muons with opposite charges), at least one light jet, and exactly one $b$-tagged jet.

\begin{figure}[t]
  \centering
  \includegraphics[width=0.85\linewidth]{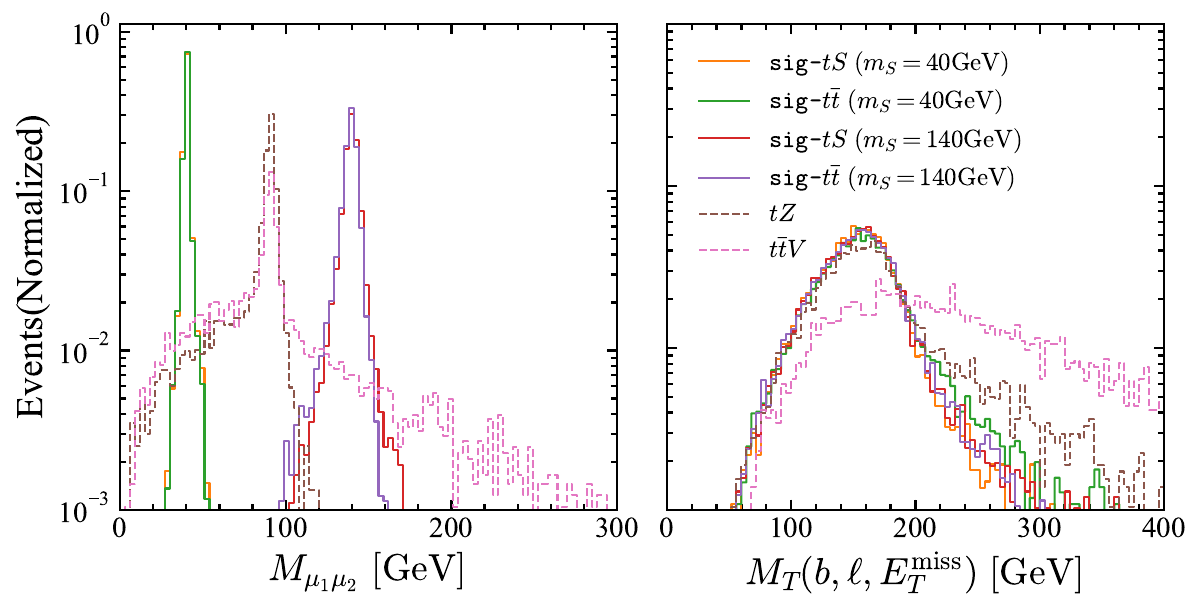}
  \caption{$M_{\mu_1\mu_2}$ (left) and $M_T(b, \ell, \slashed E_T)$ (right) distributions for the signal with two benchmark masses $m_S=40\GeV$ and $140\GeV$ as well as for the two dominant backgrounds $ ttV $ and $ tZ $, where the basic and the multiplicity cut have been imposed.}
  \label{fig:distribution}
\end{figure}

In order to further reduce the SM backgrounds, we also require a full reconstruction of the scalar singlet $ S $ and the spectator top quark. To this end, the scalar resonance $ S $ is reconstructed from the invariant mass $ M_{\mu_1\mu_2} $ of an opposite-sign muon pair (marked as $\mu_1, \mu_2$).\footnote{If an event contains more than one pair of qualified muons, the pair with an invariant mass closer to the input $ m_{S} $ will be selected. In this way, the $M_{\mu_1\mu_2}$ distributions of the backgrounds depend on the prior $m_{S}$. However, our numerical results show that the differences of the distributions caused by the different $m_{S}$ values are insignificant.} Then, the remaining lepton ($\ell$), combined with the $b$-tagged jet and the missing transverse energy, is used to reconstruct the transverse mass of the spectator top quark, with~\cite{ParticleDataGroup:2024cfk}
\begin{align}\label{eq:rec-mt}
   M_T(b, \ell, \slashed E_T) = \sqrt{\left(E_{T,b\ell} + \slashed E_T\right)^2 - \left(\vec{p}_{T,b\ell} + \vec{\slashed{p}}_T\right)^2} \,,
\end{align}
where the missing transverse momentum $ \vec{\slashed{p}}_T $ is reconstructed as the negative vector sum of the transverse momenta of all the visible final states, and the missing transverse energy is defined as $ \slashed E_T =  |\vec{\slashed{p}}_T|$. For the $b - \ell$ system, its transverse momentum and transverse energy are defined, respectively, as $\vec{p}_{T, b\ell} = \vec{p}_{T,b} + \vec{p}_{T,\ell}$ and $E_{T,b\ell} = \sqrt{|\vec{p}_{T, b\ell}|^2 + M^2_{b\ell}} $, with the invariant mass $ M^2_{b\ell} = (p_{b} + p_{\ell})^2 $. In figure~\ref{fig:distribution}, we show the distributions of $M_{\mu_1 \mu_2}$ and $M_T(b, \ell, \slashed E_T)$ for the signals with two benchmark points $m_S = 40\GeV$ and $140\GeV$ as well as for the two dominant backgrounds $ ttV $ and $ tZ $. It can be seen that the $ M_{\mu_1 \mu_2} $ distributions of the signal channels have peaks at the corresponding $ m_{S} $. However, the peaks for the backgrounds appear at $ m_Z $. In addition, as expected, the $ M_T(b, \ell,\slashed E_T) $ distributions have an end-point at $ m_t $.

According to the above analysis, we require the invariant mass $ M_{\mu_1 \mu_2} $ to lie in the regime of $ \pm 20 \GeV$ around the input $ m_{S} $, while the reconstructed transverse masses $ M_T(b, \ell, \slashed{E}_T) $ are required to satisfy a loose requirement, being less than 200$ \GeV $. The selection criterion on $ M_{\mu_1 \mu_2} $ means that we need to define the signal regions depending on the prior $m_S$ values. Finally, we summarize the cuts at the detector level as follows:
\begin{enumerate}[itemindent=1em]
\item[\textbf{cut-1:}] basic cut with $ p_T^{\ell} > 15 \GeV$, $ p_T^{j} > 25 \GeV$,  $|\eta^{\ell,j}| < 2.5 $, $ \mathrm{max}(p_T^{\ell}) > 25 \GeV$,
\item[\textbf{cut-2:}] multiplicity cut with $ N(\ell) = 3 $, $ N(\mu^+\mu^-) \ge 1 $, $ N(j) \ge 1 $, $ N(b) = 1 $,
\item[\textbf{cut-3:}] invariant mass cut with $ |M_{\mu_1\mu_2}-m_{S}| < 20 \GeV$, $ M_T(b,\ell,\slashed{E}_T) < 200 \GeV$.
\end{enumerate}
The cut flows for the signals with several benchmark values of $m_S$ and the SM backgrounds are shown in tables~\ref{tab:cutflow-3l-sig} and \ref{tab:cutflow-3l-bkg}, respectively. Here we consider an integrated luminosity of $150\fbi$ and a centre-of-mass energy of  $\sqrt{s} = 14 \TeV$.

\begin{table}[t]
  \centering
  \resizebox{\linewidth}{!}{
  \tabcolsep=0.1cm
  \renewcommand\arraystretch{1.2}
  \begin{tabular}{c | rrrrrrrr}
    \bottomrule
    $ m_{S} $ & 20$\GeV$ & 40$\GeV$ & 60$\GeV$ & 80$\GeV$ & 100$\GeV$ & 120$\GeV$ & 140$\GeV$ & 160$\GeV$ \\ \midrule
    init   & 42.15(1.72) & 38.76(1.53) & 33.47(1.30) & 26.69(1.10) & 19.13(0.92) & 11.61(0.77) &  5.20(0.64) &  1.07(0.54) \\ 
    cut-1  & 36.49(1.49) & 34.47(1.36) & 30.40(1.20) & 24.64(1.03) & 17.83(0.87) & 10.88(0.73) &  4.91(0.62) &  1.03(0.52) \\ 
    cut-2  &  4.14(0.12) &  3.98(0.11) &  3.70(0.11) &  3.18(0.10) &  2.40(0.09) &  1.42(0.08) &  0.59(0.07) &  0.12(0.06) \\ 
    cut-3  &  3.50(0.11) &  3.38(0.10) &  3.17(0.10) &  2.73(0.09) &  2.06(0.08) &  1.22(0.07) &  0.51(0.06) &  0.09(0.05) \\ 
    \midrule
    Eff.~[$$\%$$] & 8.30(6.09)& 8.72(6.71)& 9.47(7.37)& 10.22(8.19)& 10.77(8.63)& 10.54(9.00)& 9.87(9.33)& 8.76(9.43)\\
    \toprule
  \end{tabular}}
  \caption{Event yields of $t\bar{t}$ ($t S$) production channels for eight benchmark values of $m_{S}$ with $\ycts=0.001$, assuming an integrated luminosity of $150\fbi$ and a centre-of-mass energy of $\sqrt{s} = 14 \TeV$. The last line gives the accumulated efficiencies after the three cuts.}
  \label{tab:cutflow-3l-sig}
\end{table}

\begin{table}
  \centering
  \resizebox{\linewidth}{!}{
  \tabcolsep=0.1cm
  \renewcommand\arraystretch{1.2}
  \begin{tabular}{c | rrrrrrrrrrr}
    \bottomrule
    \multirow{2}{*}{cut}& \multirow{2}{*}{init} & \multirow{2}{*}{cut-1} & \multirow{2}{*}{cut-2} &\multicolumn{8}{c}{cut-3} \\
    \cline{5-12} 
                    & & & & 20$\GeV$  & 40$\GeV$  & 60$\GeV$  & 80$\GeV$  & 100$\GeV$ & 120$\GeV$ & 140$\GeV$ & 160$\GeV$\\ \midrule
        $t\bar{t}$  & 1031746 & 841756 & 126.39 &  18.06 &  30.95 &  25.79 &  23.21 &  20.63 &  20.63 &  20.63 &  20.63 \\ 
        $t\bar{t}V$ &     923 &    836 &  45.45 &   1.39 &   3.04 &   4.37 &  10.89 &  11.14 &   3.95 &   3.14 &   2.31 \\ 
        $tW$        &  595703 & 475172 &   1.99 &   0.00 &   1.99 &   1.99 &   0.00 &   0.00 &   0.00 &   0.00 &   0.00 \\ 
        $tZ$        &    2952 &   2619 & 160.11 &   5.07 &   9.82 &  15.97 & 128.01 & 129.28 &  12.91 &   6.95 &   5.07 \\ 
        $VV$        &  972368 & 673783 & 145.86 &   4.86 &  19.45 &  19.45 &  87.51 &  92.37 &   9.72 &   4.86 &   4.86 \\ 
        $VVV$       &      76 &     72 &   0.30 &   0.01 &   0.01 &   0.02 &   0.07 &   0.06 &   0.01 &   0.01 &   0.01 \\
    \toprule
  \end{tabular}}
  \caption{Event yields for different backgrounds after each cut, assuming an integrated luminosity of $150\fbi$ and a centre-of-mass energy of $\sqrt{s} = 14 \TeV$.}
  \label{tab:cutflow-3l-bkg}
\end{table}

\begin{figure}[t]
  \centering
  \includegraphics[width=\linewidth]{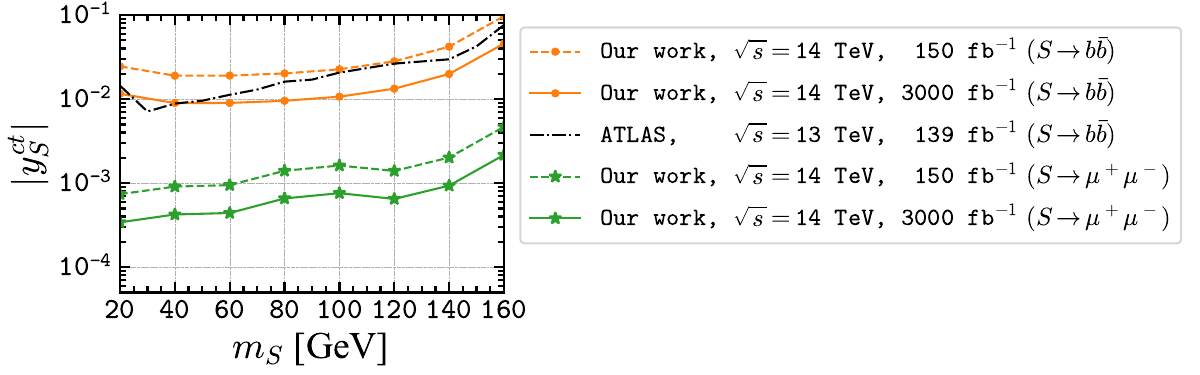}
  \caption{The expected $95\%$ CL upper bounds on $|\ycts|$ in the $\mu^+ \mu^-$ and $b \bar b$ channels, assuming $ \mB(S\to \mu^+\mu^-)=100\% $ and $ \mB(S\to b\bar{b})=100\%$, respectively. The bounds are obtained for $\mL=150\fbi$ (dashed line) and $3000 \fbi$ (solid line). The ATLAS bound from a search of $t \to q X$, with $X \to b \bar b$, produced in $t \bar t$ events based on $\mL=139 \fbi$~\cite{ATLAS:2023mcc} is also shown (gray dashed line).}
  \label{fig:limit-coupling}
\end{figure}

\begin{figure}[t]
  \centering
  \includegraphics[width=\linewidth]{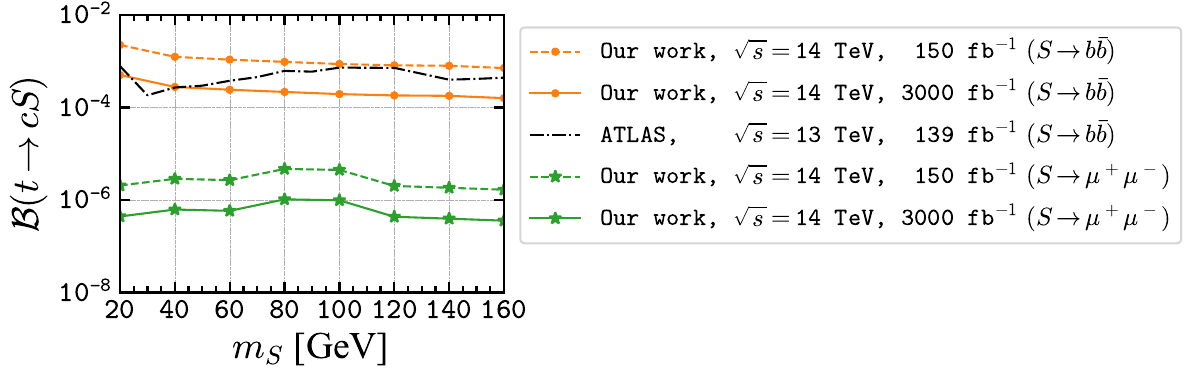}
  \caption{The expected $95\%$ CL upper bounds on the branching ratio of $t \to c S$ decay, which is derived by using the expected bounds on $|\ycts|$ shown in figure~\ref{fig:limit-coupling}. The other captions are the same as in figure~\ref{fig:limit-coupling}.}
  \label{fig:limit-br}
\end{figure}

The expected upper bounds on the parameters $m_S$ and $\ycts$ are derived by using the statistical significance $Z$ defined in eq.~(\ref{eq:z-test}), where $S$ and $B$ denote the event numbers after all the three cuts for the signals and backgrounds, respectively. For the integrated luminosities of $\mL = 150\fbi$ and $3000\fbi$, the expected $95\%$ upper bounds on $|\ycts|$ for several benchmark values of $m_S$ are shown in figure~\ref{fig:limit-coupling}. By using eq.~\eqref{eq:tdecay}, we can then derive the corresponding upper bounds on the branching ratio $\mB(t\to c S)$, which are shown in figure~\ref{fig:limit-br}. For $\mL= 150\fbi$ with $\sqrt{s}=14\TeV$, our results are consistent with the analysis for the $13\TeV$ LHC in ref.~\cite{Castro:2020sba}.

\subsection[Three-\texorpdfstring{$b$}{b}-jet final states through \texorpdfstring{$S\to b\bar{b}$}{H0 -> bb} channel]{Three-$\boldsymbol{b}$-jet final states through $\boldsymbol{S\to b\bar{b}}$ channel}
\label{sec:LHC:bb}

In this subsection, we focus on the scenario in which the scalar $ S $ decays to $ b\bar{b} $ after the production, as shown in figure~\ref{fig:diagram:LHC}. The full decay chains can be written as
\begin{align*}
    p p & \to t(\to b \ell \nu_\ell) t(\to q S, S\to b\bar{b} )  &&\to\quad \geqslant 1j + 3b + 1\ell + \slashed{E}_T, \\
    p p & \to t(\to b \ell \nu_\ell) S(\to b\bar{b}) && \to\quad \geqslant 0j + 3b + 1\ell + \slashed{E}_T.
\end{align*}
The signals are characterized by at least three jets (three of them are $b$-tagged), exactly one isolated lepton, and missing transverse energy. 

For this channel, the dominant background is expected to come from $ t\bar{t} $ production associated with two additional jets. Due to the potential misidentification of light jets as the $ b $-tagged jets, the single top and the single gauge boson production with two $ b$-jets are also included. All the backgrounds considered are summarized as
\begin{align*}
    t\bar{t}jj: && pp &\to t (\to bjj ) \bar{t}(\to \bar{b} \ell \bar{\nu}_\ell) &&\to\quad \geqslant 2j + 2b + 1\ell + \slashed{E}_T, \\
    t\bar{t}bb: && pp & \to t(\to bjj ) \bar{t}(\to \bar{b} \ell \bar{\nu}_\ell) +bb && \to\quad \geqslant 2j + 4b + 1\ell + \slashed{E}_T, \\ 
    tW: && pp &\to t (\to bjj ) W (\to \ell\nu_\ell) && \to \quad \geqslant 2j + b + 1\ell + \slashed{E}_T, \\
    tb~({\textrm{s-channel}}): && pp &\to t (\to b\ell\nu_\ell ) + b && \to \quad \geqslant 0j + 2b + 1\ell + \slashed{E}_T, \\
    tb~({\textrm{t-channel}}): && pp &\to t (\to b\ell\nu_\ell ) + j && \to \quad \geqslant 1j + 1b + 1\ell + \slashed{E}_T, \\
    Wbb: && pp &\to W (\to \ell\nu_\ell) + bb &&\to\quad 2b + 1\ell + \slashed{E}_T, \\
    Zbb: && pp &\to Z (\to \ell\ell) + bb &&\to\quad 2b + 2\ell.  
\end{align*}
For the $ t\bar{t}jj $, $ Wbb $ and $ Zbb $ channels, the samples are merged up to two extra light jets, while the samples of the $ tW $ channel are merged up to one extra light jet.

Both signal and background events are simulated and analyzed by using the same method as in the last subsection. Furthermore, at the detector level, we require the jets to have $p^j_T>25\GeV$ and $|\eta^j|<2.5$, and the leptons to have $p^\ell_T>15\GeV$ and $|\eta^\ell|<2.5$. The events are also required to fulfill the following selection criteria:
\begin{enumerate}[itemindent=1em]
    \item[\textbf{cut-1:}] $N(\ell)=1$, i.e., exactly one isolated lepton ($e^\pm$ or $\mu^\pm$),
    \item[\textbf{cut-2:}] $N(b)=3$, i.e., exactly three $b$-tagged jets,
    \item[\textbf{cut-3:}] $N(j)=1$ or $2$, i.e., one or two additional light jets. 
\end{enumerate}
For an integrated luminosity of $150\fbi$, the above cut flows for the signals with several benchmark values of $m_S$ and the backgrounds are shown in tables~\ref{tab:3bcutflow-sig} and \ref{tab:3bcutflow-bkg}, respectively.

\begin{table}[t]
  \centering
  \resizebox{\linewidth}{!}{
  \tabcolsep=0.1cm
  \renewcommand\arraystretch{1.2}
  \begin{tabular}{c|rrrrrrrr}
    \bottomrule
    $m_{S}$  & 20$\GeV$ & 40$\GeV$ & 60$\GeV$ & 80$\GeV$ & 100$\GeV$ & 120$\GeV$ & 140$\GeV$ & 160$\GeV$ \\ \midrule
    init   & 42.51(1.74) & 38.92(1.53) & 33.24(1.30) & 26.15(1.09) & 18.45(0.91) & 11.10(0.76) &  5.07(0.63) &  1.30(0.53) \\ 
    cut-1  & 23.71(1.02) & 21.48(0.89) & 18.24(0.75) & 14.34(0.62) & 10.11(0.52) &  6.06(0.43) &  2.78(0.35) &  0.71(0.30) \\ 
    cut-2  &  1.55(0.02) &  2.28(0.05) &  2.24(0.07) &  1.94(0.07) &  1.52(0.07) &  0.97(0.06) &  0.46(0.06) &  0.12(0.05) \\ 
    cut-3  &  0.87(0.01) &  1.44(0.03) &  1.42(0.03) &  1.24(0.04) &  0.97(0.04) &  0.62(0.03) &  0.28(0.03) &  0.07(0.03) \\ 
    \midrule 
    Eff. [\%] & 2.06(0.64)& 3.71(1.65)& 4.26(2.64)& 4.75(3.32)& 5.27(3.92)& 5.59(4.56)& 5.59(4.97)& 5.28(5.41)\\ 
    \toprule
  \end{tabular}}
  \caption{Same as in table~\ref{tab:cutflow-3l-sig}, but for the signal channel with three-$b$-jet final states.}
  \label{tab:3bcutflow-sig}
\end{table}

\begin{table}[t]
  \centering
  \resizebox{\linewidth}{!}{
  \tabcolsep=0.1cm
  \renewcommand\arraystretch{1.2}
  \begin{tabular}{c|rrrrrrr}
    \bottomrule
      & $t\bar{t}jj$ & $t\bar{t}bb$ &  $Wbb$ &      $tW$ &     $Zbb$ & $tb$ (t-channel) & $tb$ (s-channel) \\ \midrule
      init   &  34520720 &    705553 &  42821826 &  10865010 &  12762239 &  31941000 &   1511010 \\ 
      cut-1  &  18448881 &    377605 &  20204002 &   2597650 &   5573320 &   4483686 &    185839 \\ 
      cut-2  &   3543108 &    102256 &    493736 &    163866 &    117192 &    107769 &      8398 \\ 
      cut-3  &   1782085 &     58575 &    329158 &    101392 &     65298 &     50275 &      3512 \\
    \toprule
  \end{tabular}}
  \caption{Event yields for the backgrounds after each cut, assuming $\mL=150\fbi$.}
  \label{tab:3bcutflow-bkg}
\end{table}

As in the last subsection, these results can be interpreted as the expected bounds on the couplings and mass of the scalar singlet $S$. For integrated luminosities of $\mL=150\fbi$ and $3000\fbi$ at the LHC with $\sqrt{s} = 14\TeV$, the expected $95\%$ upper bounds on $|\ycts|$ for several benchmark values of $m_S$ are shown in figure~\ref{fig:limit-coupling}, and the corresponding upper limits on $\mB( t \to c S)$ are shown in figure \ref{fig:limit-br}. For comparison, the ATLAS bound from the search of $t\to q X$, with $X \to b \bar b$, produced in $t \bar t$ events is also shown, which is based on the Run 2 dataset corresponding to an integrated luminosity of $139\fbi$ at $\sqrt{s} = 13 \TeV$~\cite{ATLAS:2023mcc}. For $\mL= 3000\fbi$ with $\sqrt{s}=14\TeV$, our results are consistent with the analysis in ref.~\cite{Banerjee:2018fsx}.

\subsection{Multiple final states}
\label{sec:LHC:combined}

\begin{figure}[t]
  \centering
  \includegraphics[width=0.85\linewidth]{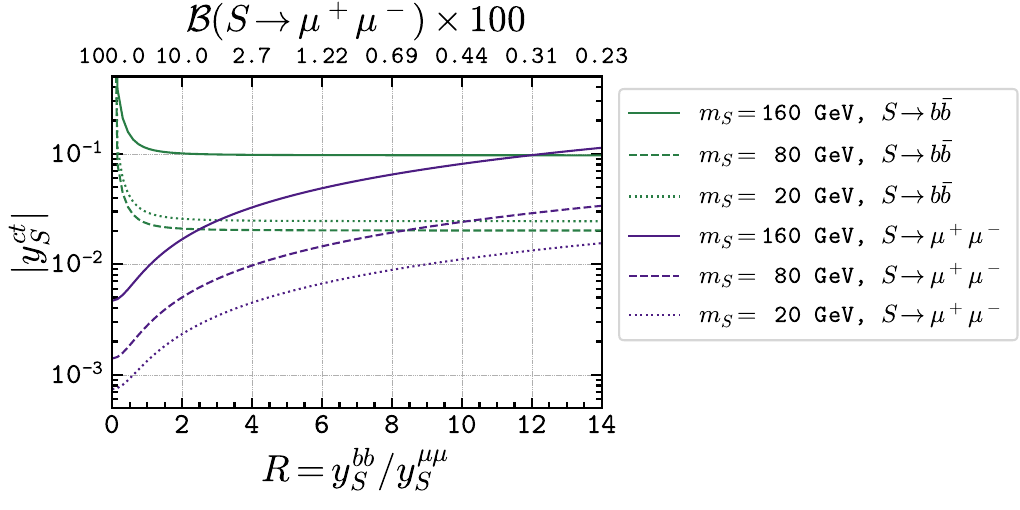}
  \caption{The expected $95\%$ CL upper bounds on $|\ycts|$ as a function of $R=\ybbs/\ymms$ for $m_S = 20$, $80$ and $160\GeV$. The green and purple curves correspond to the bounds obtained in the $b \bar b$ and $\mu^+ \mu^-$ channels with $ \mL = 150\fbi$, respectively. The branching ratios $\mB(S \to \mu^+ \mu^-)$ for different values of $R$ are also given in the upper axis.}
  \label{fig:limit-R}
\end{figure}

The discussions in the last two subsections assume that only one decay channel of the scalar singlet $S$ dominates each time. However, the $b \bar b$ and $\mu^+ \mu^-$ decay channels could exist simultaneously. Then, according to eq.~\eqref{eq:h0decay}, the branching ratio of the $\mu^+ \mu^-$ channel can be written as
\begin{align}\label{eq:br-h0}
  \mB (S\to\mu^+\mu^-) &= \frac{\beta_\mu^3}{\beta_\mu^3+3R^2 \beta_b^3},
\end{align}
with the ratio $R\equiv \ybbs/\ymms$, and $\mB(S \to b \bar b)=1-\mB(S \to \mu^+ \mu^-)$. Therefore, the sensitivities to the $tcS$ couplings from the LHC direct searches depend on the ratio $R$. In figure~\ref{fig:limit-R}, the expected upper bounds on $|\ycts|$ from the $\mu^+\mu^-$ and $b \bar b$ channels are shown as a function of $R$. For $R \lesssim 10$, we can see that the expected upper bounds from the $\mu^+ \mu^-$ channel are stronger than from the $b \bar b$ channel. In figure~\ref{fig:limit-R}, we also show the resulting branching ratios $\mB(S \to \mu^+ \mu^-)$ for different values of $R$.\footnote{For $m_S > 20\GeV$, the branching ratio $\mB(S \to \mu^+ \mu^-)$ shows a very weak dependence on the scalar mass $m_S$. Therefore, $m_S=20\GeV$ is taken in the evaluation of $\mB(S \to \mu^+ \mu^-)$.} In a more realistic situation, the scalar singlet $S$ can even have other decay channels, such as the $\gamma\gamma$ and $\tau^+ \tau^-$ modes. In this case, considering the $\mu^+ \mu^-$ channel, the expected bound for a given value of $\mB(S \to \mu^+ \mu^-)$ is equivalent to that for the corresponding $R$ value shown in figure~\ref{fig:limit-R}. For example, the bounds obtained by assuming $\mB(S \to \mu^+ \mu^-)=100\%$, $31\%$ and $0.44\%$ are equivalent to the results for $R=0$, $1$ and $10$ in figure~\ref{fig:limit-R}, respectively.

\section{Combined analysis}
\label{sec:combined analysis}

In this section, we perform a combined analysis under the constraints derived in the previous sections. The $tcS$ couplings are assumed to be CP-conserving and CP-violating in the following two subsections, respectively.

\subsection[CP-conserving \texorpdfstring{${tcS}$}{tcS} couplings]{CP-conserving $\boldsymbol{tcS}$ couplings}
\label{sec:combined analysis:CPC}

With our specification for the couplings of the scalar singlet $S$ with the SM fermions (cf. eqs.~(\ref{eq:L:tcS}) and \eqref{eq:L:ffS}), there involve totally six free coupling parameters $\ycts$, $\yctp$, $\ymms$, $\ymmp$, $\ybbs$, and $\ybbp$. Since all the observables considered are not sensitive to the difference between $\ybbs$ and $\ybbp$, we take $\ybbp=0$ without loss of generality. For the $\Bsmumu$ decay, the NP contribution involves only the combination $\ycts+i\yctp \equiv \yctr$ (cf. eq.~(\ref{eq:matching-wcs})), which is real for the CP-conserving $tcS$ interaction. Therefore, it is convenient to use the chiral basis, i.e., $\yctl$ and $\yctr$. Since these two couplings are added quadratically in the branching ratio of $t \to c S$ decay (cf. eq.~\eqref{eq:tdecay}), the upper bound on $|\yctr|$ from the LHC direct searches can be obtained by setting $\yctl=0$ and vice versa. Therefore, we are left with only the four coupling parameters $\yctr$, $\ymms$, $\ymmp$, and $\ybbs$. For the first three ones, we consider the following two scenarios:
\begin{align}\label{eq:scenario:CPC}
  \big( \yctr,\, \ymms,\, \ymmp \big)=
  \begin{cases}
    \big(  \times  ,\,     \times  ,\,     \;0  \big) & \text{scenario \scenario{S}{1},} \\
    \big(  \times  ,\,        \;0  ,\,  \times  \big) & \text{scenario \scenario{P}{1},}
  \end{cases}
\end{align}
where $\times$ denotes a nonzero entry. For each scenario, several typical benchmark values of $\ybbs/y_{S,P}^{\mu\mu}$ and $m_S$ will be considered. It is noted that CP is conserved in both the quark and lepton sectors in scenario~\scenario{S}{1}, but violated in the lepton sector in scenario~\scenario{P}{1}.

\begin{figure}[t]
  \centering
  \includegraphics[width=\linewidth]{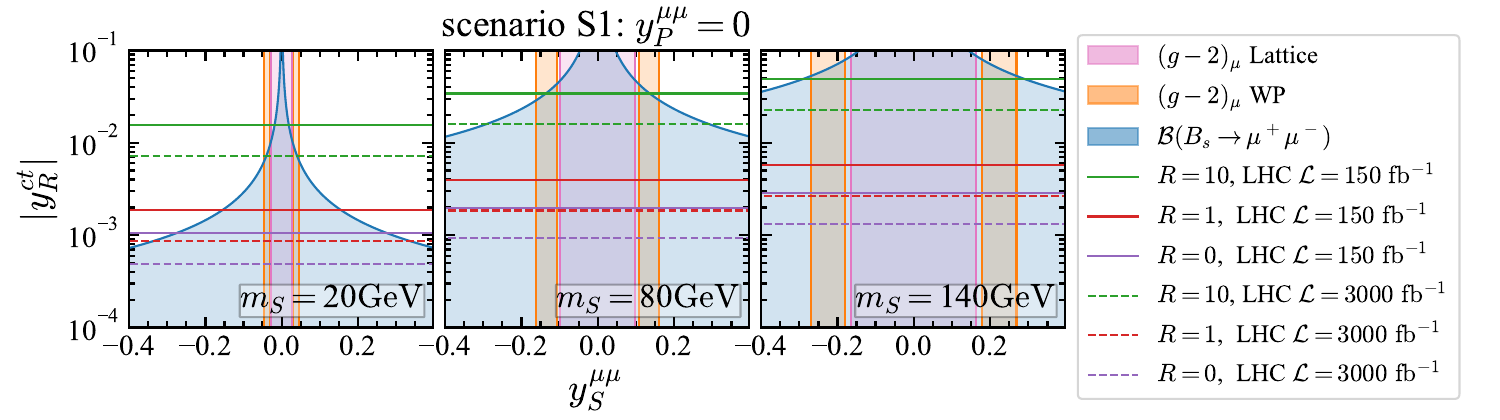}
  \\[1em]
  \includegraphics[width=\linewidth]{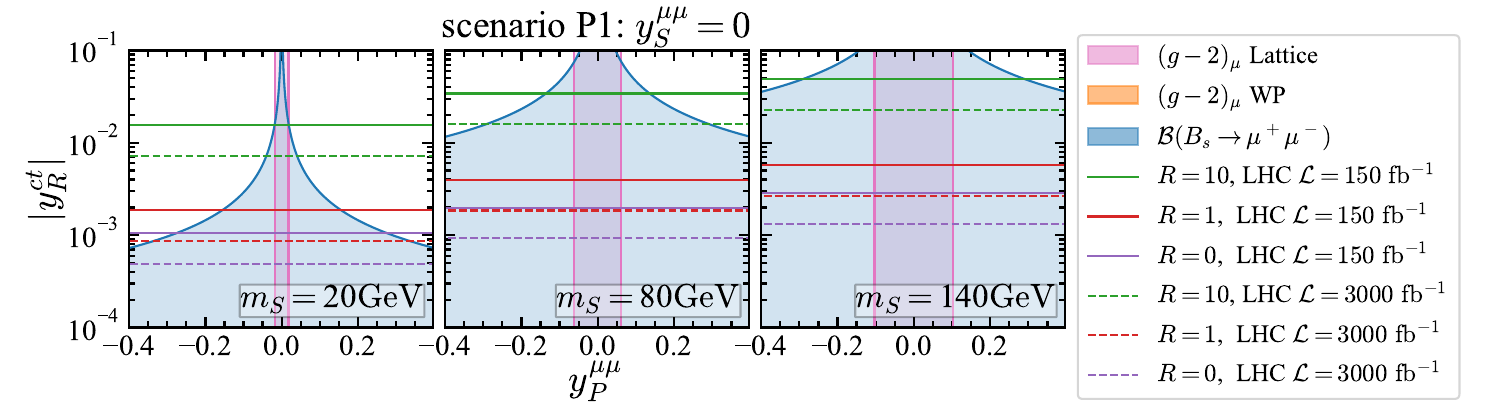}
  \caption{Constraints on the couplings $(\yctr,\, \ymms)$ in scenario~\scenario{S}{1} and on $(\yctr,\, \ymmp)$ in scenario~\scenario{P}{1}. The $2\sigma$ allowed regions by the branching ratio $\BR$ and the $(g-2)_\mu$ from eq.~(\ref{eq:muong-2}) (eq.~\eqref{eq:muong-2:2}) are shown by blue and orange (pink), respectively. For each $R=\ybbs/y_{S,P}^{\mu\mu}$ value, the expected $95\%$ CL upper bounds at the LHC with $150\fbi$ and $3000\fbi$ are shown by the horizontal lines.}
  \label{fig:combined:CPC}
\end{figure}

In each scenario, we consider the constraints from the $\Bsmumu$ decay, $(g-2)_\mu$, and the direct searches at the LHC with the decay channel $S \to \mu^+ \mu^-$. For the benchmark masses $m_S=20$, $80$ and $140\GeV$, the allowed parameter spaces of $(\yctr,\,y_{S,P}^{\mu\mu})$ are shown in figure~\ref{fig:combined:CPC}. From this figure, we make the following observations:
\begin{itemize}
\item In scenario \scenario{S}{1}, considering the SM prediction from the Muon $g-2$ Theory Initiative~\cite{Aoyama:2020ynm}, the $(g-2)_\mu$ can be explained by a nonzero scalar coupling $\ymms$. In figure~\ref{fig:combined:CPC}, we also show the constraints by considering the recent lattice result for the HVP contribution in ref.~\cite{Boccaletti:2024guq}. In the allowed parameter regions of $\ymms$ corresponding to the two SM predictions of $(g-2)_\mu$, we can see that the upper bounds from the $\Bsmumu$ decay are weaker than the expected $95\%$ CL upper bounds at the LHC for $R \lesssim 5$.

\item In scenario \scenario{P}{1}, the main difference from scenario \scenario{S}{1} is the constraint from $(g-2)_\mu$. Considering the SM prediction from the Muon $g-2$ Theory Initiative~\cite{Aoyama:2020ynm}, the $(g-2)_\mu$ cannot be explained by the coupling $\ymmp$ in this scenario. Therefore, we show in figure~\ref{fig:combined:CPC} only the constraints from the recent lattice result for the HVP contribution in ref.~\cite{Boccaletti:2024guq}. For the $\Bsmumu$ decay, the constraints in this scenario are almost identical to the ones in scenario \scenario{S}{1}, which can be understood from the semi-numerical result in eq.~(\ref{eq:Bsmumu:simi-num}), keeping in mind that $\yctr$ is real for CP-conserving $tcS$ interaction. In addition, we find that the allowed regions by both the $\Bsmumu$ and $(g-2)_\mu$ are weaker than the expected $95\%$ CL upper bounds at the LHC for $R \lesssim 10$.

\item For these two scenarios, the upper bounds on the $tcS$ couplings from the $\Bsmumu$ decay relative to the expected bounds at the LHC become stronger for lighter scalar $S$.
\end{itemize}

\begin{figure}
  \centering
  \includegraphics[width=0.45\linewidth]{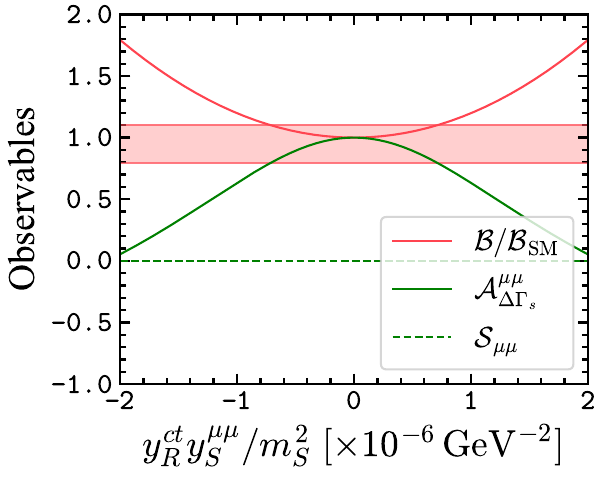}\;
  \includegraphics[width=0.45\linewidth]{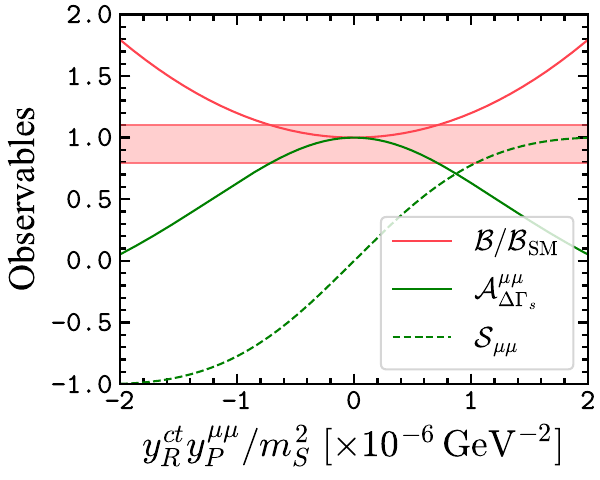}
  \caption{The predicted $\mB/\mB_{\text{SM}}$, $\ADG$ and $\Smumu$ of the $\Bsmumu$ decay as a function of $\yctr \ymms/m_S^2$ in scenario \scenario{S}{1} (left) and $\yctr \ymmp/m_S^2$ in scenario \scenario{P}{1} (right). The red band denotes the $2\sigma$ range of $\mB_{\text{exp}}/\mB_{\text{SM}}$ obtained from eqs.~\eqref{eq:br_bsmumu:exp} and \eqref{eq:br_bsmumu:SM}.}
  \label{fig:prediction:CPC}
\end{figure}

For the $\Bsmumu$ decay, the NP contributions are controlled, respectively, by the products $\yctr \ymms/m_S^2$ in scenario \scenario{S}{1} and by $\yctr \ymmp/m_S^2$ in scenario \scenario{P}{1}. The predicted branching ratio and CP observables of the decay are shown as a function of these products in figure~\ref{fig:prediction:CPC}. We can see that both $\BR$ and $\ADG$ are almost identical for the two scenarios. However, the CP-violating $\mu\mu S$ interaction in scenario \scenario{P}{1} can induce a nonzero phase in the Wilson coefficient $\mC_P$, making $\Smumu$ deviate from the SM value of zero. Therefore, we can use the CP observable $\Smumu$ to differentiate between these two scenarios.

\subsection[CP-violating \texorpdfstring{${tcS}$}{tcS} couplings]{CP-violating $\boldsymbol{tcS}$ couplings}
\label{sec:combined analysis:CPV}

For the $tcS$ interactions given by eq.~\eqref{eq:L:tcS}, CP violation originates from the nonzero phases of the couplings $\yctl$ and $\yctr$, i.e., $\theta_L \neq 0$ and/or $\theta_R \neq 0$, with the definition $\yctl = |\yctl|e^{i \theta_L}$ and $\yctr = |\yctr|e^{i\theta_R}$.\footnote{Alternatively, one can also use $\ycts$ and $\yctp$ as a basis to describe the $tcS$ interactions. Numerical analysis in this basis can be found in appendix~\ref{sec:combined analysis:SP basis}.} As in the last subsection, we can still choose the four coupling parameters $\yctr$, $\ymms$, $\ymmp$ and $\ybbs$, and consider the two scenarios specified by eq.~(\ref{eq:scenario:CPC}) to investigate the $\Bsmumu$ decay, $(g-2)_\mu$ and the LHC direct searches with the decay channel $S \to \mu^+ \mu^-$, but with a complex $\yctr$ coupling. 

\begin{figure}[t]
  \centering
  \includegraphics[width=\textwidth]{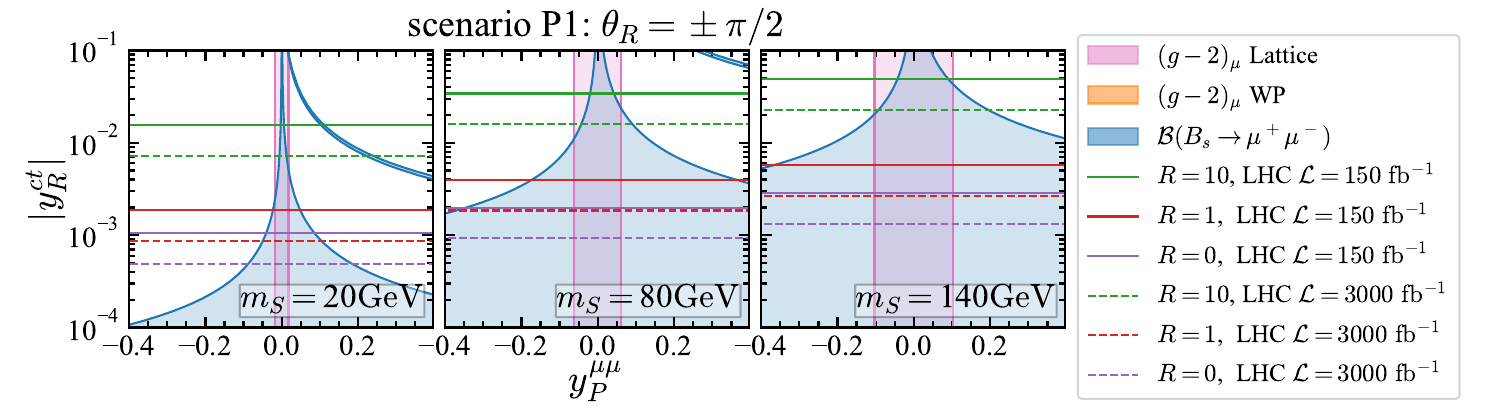}
  \caption{Constraints on the couplings $(|\yctr|,\, \ymmp)$ in scenario \scenario{P}{1} with $\yctr=\pm i |\yctr|$. The other captions are the same as in figure \ref{fig:combined:CPC}.}
  \label{fig:combined:CPV}
\end{figure}

In scenario \scenario{S}{1} with a complex $\yctr$, all the constraints remain the same as in figure~\ref{fig:combined:CPC}, because all the observables considered do not depend on the phase $\theta_R$ in this scenario. In scenario \scenario{P}{1}, on the other hand, the NP contribution to $\mC_P$ with a complex $\yctr$ can interfere with the SM contribution $\mC_{10}^\SM$, making the constraints from the $\Bsmumu$ decay different from the ones shown in figure~\ref{fig:combined:CPC}. In this case, bounds on $|\yctr|$ from the branching ratio $\BR$ depend on the phase $\theta_R$, as can be seen from eq.~\eqref{eq:Bsmumu:simi-num}. In the limit of $\theta_R \to 0$ or $\pm \pi$, the bounds on $|\yctr|$ remain the same as in figure~\ref{fig:combined:CPC}. In the limit of $\theta_R \to \pm\pi/2$, i.e., with a maximal magnitude of CP violation, the bounds are shown in figure~\ref{fig:combined:CPV}. For other $\theta_R$ values, the bounds on $|\yctr|$ lie between the ones shown in figures~\ref{fig:combined:CPC} and \ref{fig:combined:CPV}. From figure~\ref{fig:combined:CPV}, we can see that the constraints from $(g-2)_\mu$ and the direct searches at the LHC remain the same as in scenario \scenario{P}{1} with a real $\yctr$ coupling. For the $\Bsmumu$ decay, the interference between $\mC_P$ and $\mC_{10}^\SM$ makes the bounds on $|\yctr|$ much more stringent than the ones in scenario \scenario{P}{1} with a real $\yctr$ coupling. Compared to the other scenarios, in the wider region of the $\mu\mu S$ couplings, the bounds on the $tcS$ coupling are stronger than the expected $95\%$ CL upper bounds at the LHC. In particular, for $m_S=20\GeV$, the upper bounds on $|\yctr|$ in most of the $\ymmp$ regions allowed by $(g-2)_\mu$ are stronger than the expected $95\%$ CL upper bounds at the LHC for $R=10$.

\begin{figure}
  \centering
  \includegraphics[width=\linewidth]{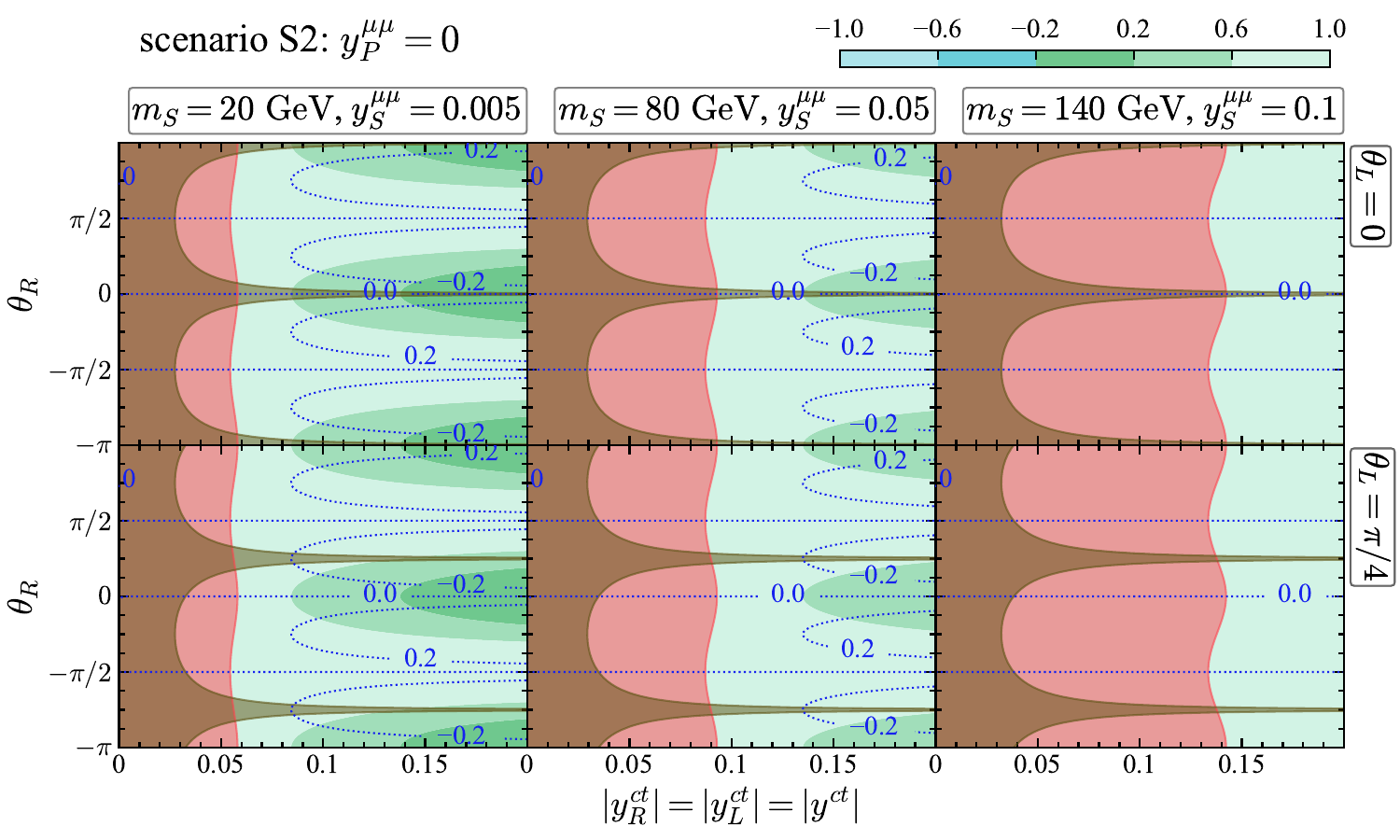}
  \\[0.5em]
  \includegraphics[width=\linewidth]{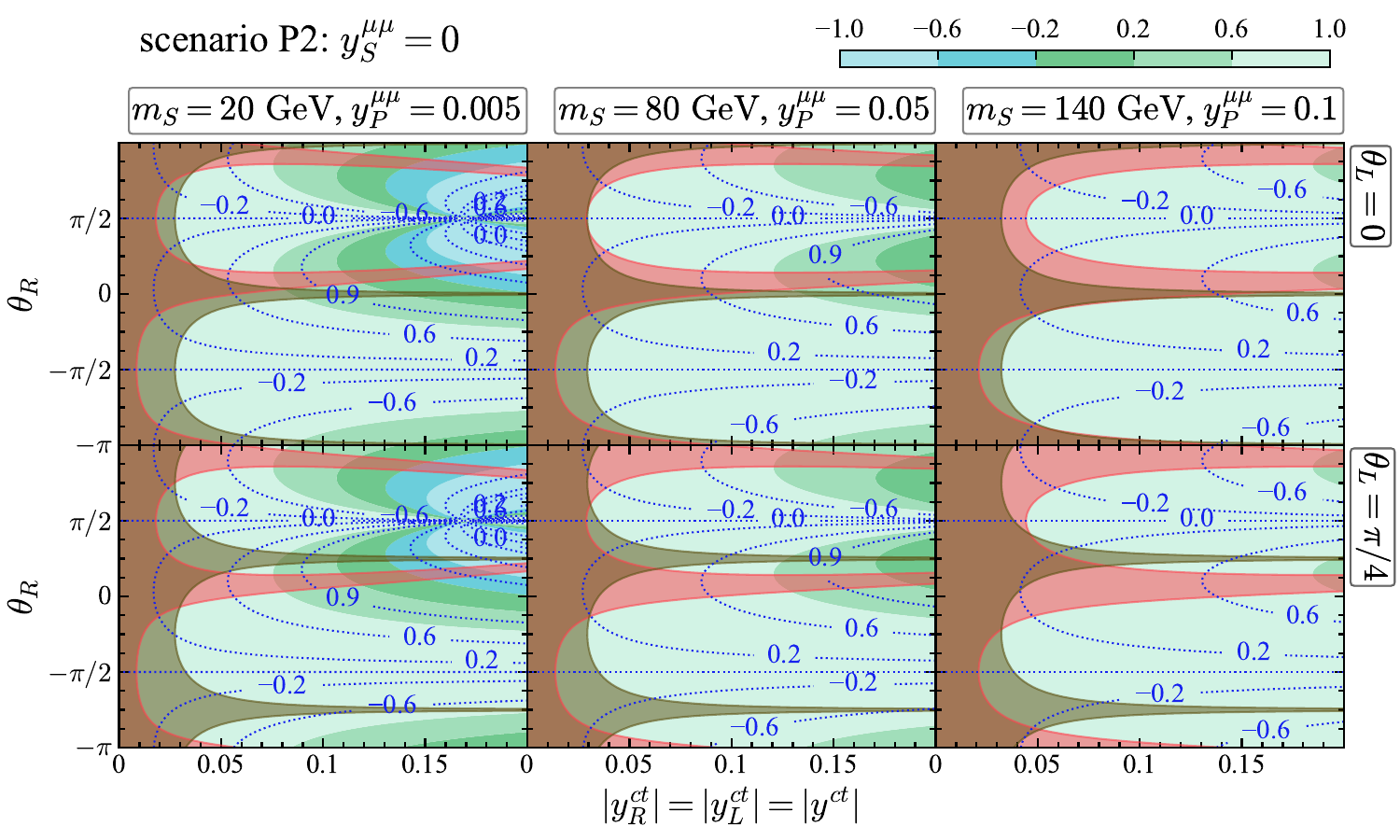}
  \caption{Constraints on $(|y^{ct}|,\, \theta_R)$ in scenarios \scenario{S}{2} (top) and \scenario{P}{2} (bottom). The $2\sigma$ allowed regions from $\BR$ and the $90\%$ CL allowed regions from the neutron EDM are shown in red and gray, respectively. The predicted $\ADG$ and $\Smumu$ are given by the colourmap and the blue curves, respectively. In the second row of each panel, results for $\theta_L=\pi/4$ are shown to illustrate the effect of a nonzero $\theta_L$.}
  \label{fig:complex:LR}
\end{figure}

The complex couplings $y_{L,R}^{ct}$ can also affect the CP observables $\ADG$ and $\Smumu$ of the $\Bsmumu$ decay, as well as the neutron EDM. In this case, the relevant free parameters include $|\yctl|$, $\theta_L$, $|\yctr|$, $\theta_R$, $\ymms$, and $\ymmp$. In order to investigate the constraints from these CP observables, we consider the following four scenarios:
\begin{align}
  \big( \yctr,\, \theta_R,\, \yctl,\, \ymms,\, \ymmp \big)=
  \begin{cases}
    \big(  |y^{ct}|        ,\;\;    \times  ,\, |y^{ct}| ,\;\; v  ,\;\; 0  \big) & \text{scenario \scenario{S}{2},} \\
    \big(  \;\;\,\,\times ,\;\;\; v       ,\, \;\;\,\,\times ,\;\; v  ,\;\;0  \big) & \text{scenario \scenario{S}{3},} \\
    \big(  |y^{ct}|        ,\;\;    \times  ,\, |y^{ct}| ,\;\; 0  ,\;\; v  \big) & \text{scenario \scenario{P}{2},} \\
    \big(  \;\;\,\,\times ,\;\;\; v       ,\, \;\;\,\,\times ,\;\; 0  ,\;\;v  \big) & \text{scenario \scenario{P}{3},} \\
  \end{cases}
\end{align}
where $\times$ denotes a nonzero entry and $v$ indicates a choice from some benchmark values. Since the $\Bsmumu$ decay involves only $\yctr=|\yctr|e^{i\theta_R}$, while the neutron EDM depends only on $\operatorname{Im}(\yctr \yctl^*)=|\yctr\yctl|\sin(\theta_R-\theta_L)$, a nonzero $\theta_L$ shifts only the neutron EDM constraints in the $\theta_R$ direction. Therefore, for simplicity, we take $\theta_L=0$ in the above four scenarios.\footnote{In order to illustrate the effect of a nonzero $\theta_L$, we show in figure~\ref{fig:complex:LR} the results with $\theta_L= \pi /4$.} It is noted that CP in the lepton sector is conserved in the scenarios \scenario{S}{2} and \scenario{S}{3}, but violated in the scenarios \scenario{P}{2} and \scenario{P}{3} by the pseudoscalar coupling $\ymmp$.

In scenarios \scenario{S}{2} and \scenario{P}{2}, where $|\yctl| = |\yctr|=|y^{ct}|$, the benchmark values $0.005$, $0.05$ and $0.1$ of $\ymms$ (for \scenario{S}{2}) and $\ymmp$ (\scenario{P}{2}) are taken for $m_S=20$, $80$ and $140\GeV$, respectively, all of which satisfy the constraints shown in figures \ref{fig:combined:CPC} and \ref{fig:combined:CPV}. Considering the $\Bsmumu$ decay and the neutron EDM, constraints on the parameters $(|y^{ct}|,\, \theta_R)$ are shown in figure \ref{fig:complex:LR}. The predicted $\ADG$ and $\Smumu$ are also given in figure \ref{fig:complex:LR}. From this figure, we make the following observations:
\begin{itemize}
  
\item After combining the constraints from $\BR$ and the neutron EDM, the parameter regions with large $|y^{ct}|$ are already excluded. In particular, the neutron EDM excludes the possibility of large NP effect on the $\Bsmumu$ decay in scenario \scenario{P}{2}.
  
\item Except in the region of $ \theta_R \approx \theta_L $ or $ \theta_R \approx \theta_L \pm \pi $, the neutron EDM constraints are compatible with the ones from $\BR$. However, the latter constraints depend on the $\mu\mu S$ couplings. For example, when the $\ymms$ and $\ymmp$ values are smaller than those used in this figure, the constraints from $\BR$ become looser and can even be weaker than from the neutron EDM.
  
\item In scenario \scenario{S}{2}, the upper bounds from $\BR$ still show a weak dependence on the phase $\theta_R$, which mainly results from the effect of $\ADG$.

\item After considering the constraint from $\BR$ and the neutron EDM, the two CP observables $\ADG$ and $\Smumu$ cannot deviate from their corresponding SM predictions by around $\pm 0.2$ in scenario \scenario{S}{2} and $\pm 0.6$ in scenario \scenario{P}{2}, respectively.
  
\item Results for $\theta_L = \pi/4$ are also given in this figure. As expected, the neutron EDM constraints are globally shifted by $\pi/4$ in the $\theta_R$ direction, while all the observables of the $\Bsmumu$ decay remain unchanged.
\end{itemize}

\begin{figure}
  \centering
  \includegraphics[width=\linewidth]{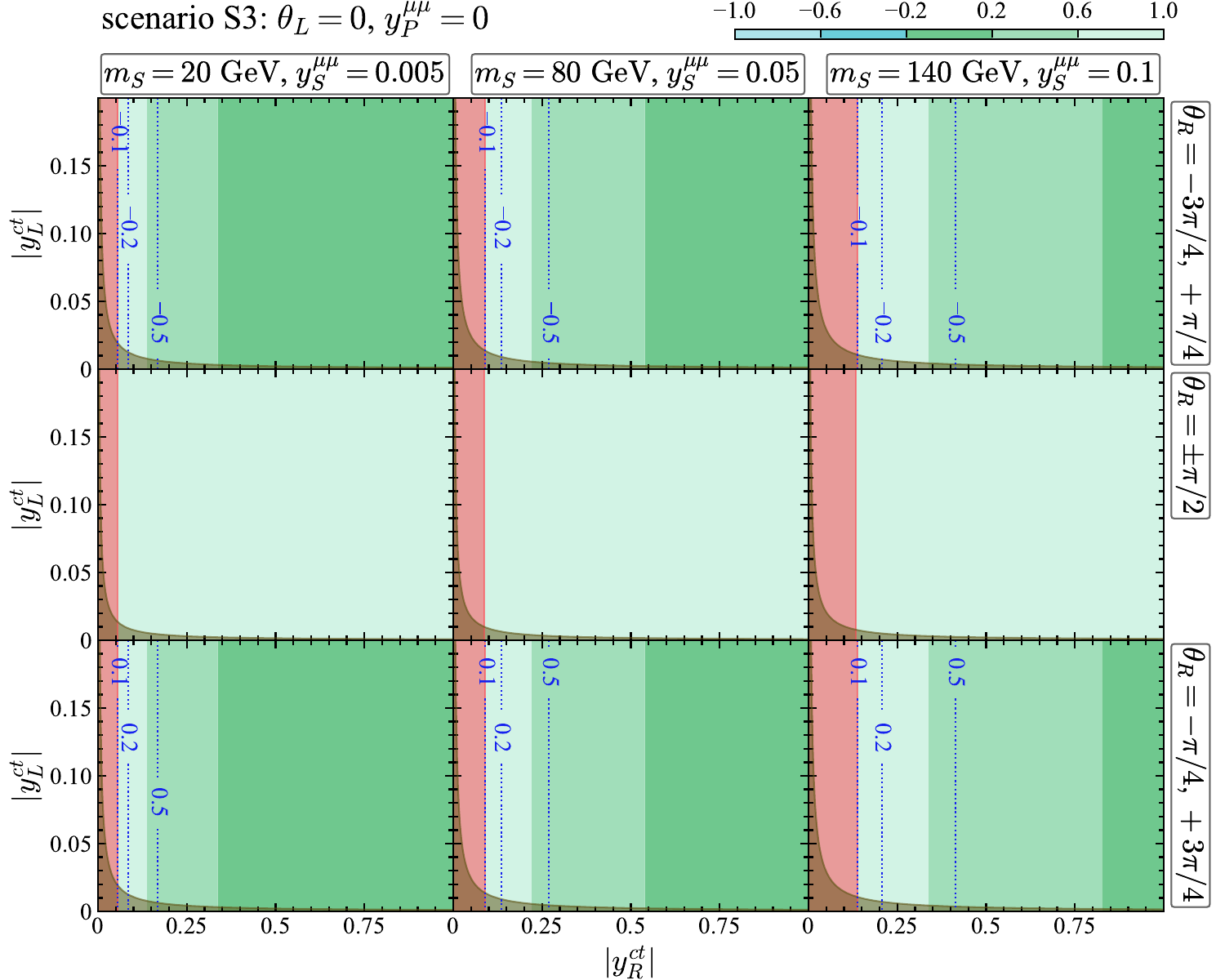}
  \caption{Constraints on $(|\yctr|,\, |\yctl|)$ in scenario \scenario{S}{3} for $\theta_R=\pm \pi/4$, $\pm\pi/2$ and $\pm 3\pi/4$. Here $\ymms=0.005$ (left), $0.05$ (middle) and $0.1$ (right) are taken for $m_S=20$, 80 and $140\GeV$, respectively. The other captions are the same as in figure~\ref{fig:complex:LR}.}
  \label{fig:complex:LR:1}
\end{figure}

\begin{figure}
  \centering
  \includegraphics[width=\linewidth]{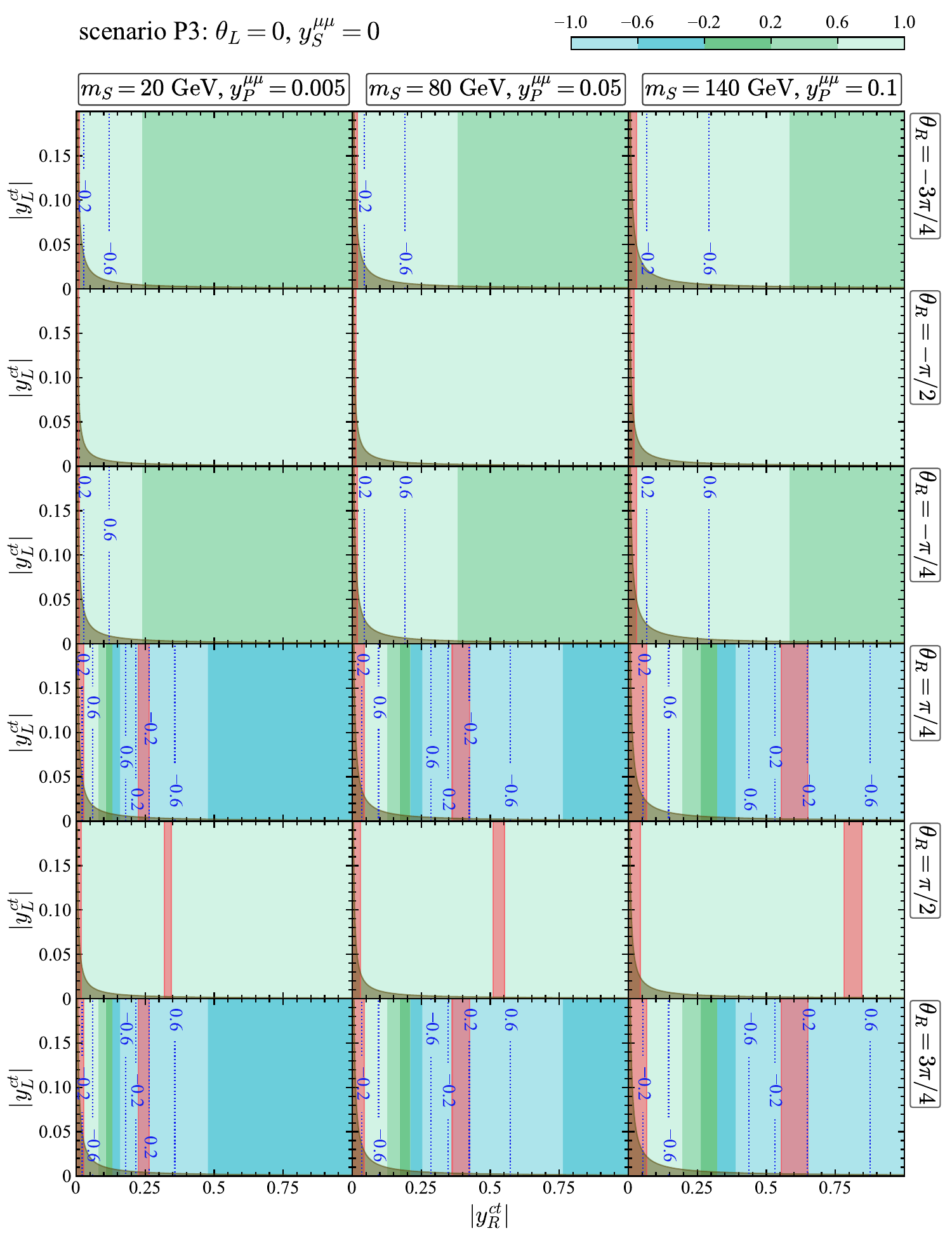}
  \caption{Constraints on $(|\yctr|,\, |\yctl|)$ in scenario \scenario{P}{3} for $\theta_R=\pm \pi/4$, $\pm\pi/2$ and $\pm 3\pi/4$.  Here $\ymmp=0.005$ (left), $0.05$ (middle) and $0.1$ (right) are taken for $m_S=20$, 80 and $140\GeV$, respectively.  The other captions are the same as in figure~\ref{fig:complex:LR}.}
  \label{fig:complex:LR:2}
\end{figure}

In scenarios \scenario{S}{3} and \scenario{P}{3}, where $|\yctl| \neq |\yctr|$, we consider the benchmark phases $\theta_R = \pm \pi/4,\,  \pm\pi/2,\, \pm 3\pi/4$.\footnote{The cases of $\theta_{R}=0$ and $\pm\pi$ respect the CP symmetry in the quark sector and have already been discussed in subsection~\ref{sec:combined analysis:CPC}.} Constraints on the parameter space in these two scenarios are shown in figures~\ref{fig:complex:LR:1} and \ref{fig:complex:LR:2}, respectively. In the region with small $|\yctl|$ (e.g., smaller than $0.02$), the branching ratio $\BR$ provides stronger upper bounds on $|\yctr|$, while the neutron EDM bounds dominate in the region with large $|\yctl|$ (e.g., larger than $0.05$). Within the theoretical framework given in subsection~\ref{sec:bsmumu}, it can be proven that both $\BR$ and $\ADG$ remain unchanged under $\theta_R \to \theta_R \pm \pi$ in scenario \scenario{S}{3} and $\theta_R \to - \theta_R \pm \pi $ in scenario \scenario{P}{3}, which can also be seen from figures~\ref{fig:complex:LR:1} and \ref{fig:complex:LR:2}, respectively. In scenario \scenario{P}{3}, this twofold ambiguity can be resolved by the observable $\Smumu$, since its sign is flipped under $\theta_R \to - \theta_R \pm \pi $. However, in scenario \scenario{S}{3}, we end up with the degeneracy even after considering $\Smumu$.

\begin{figure}[t]
  \centering
  \includegraphics[width=0.48\linewidth]{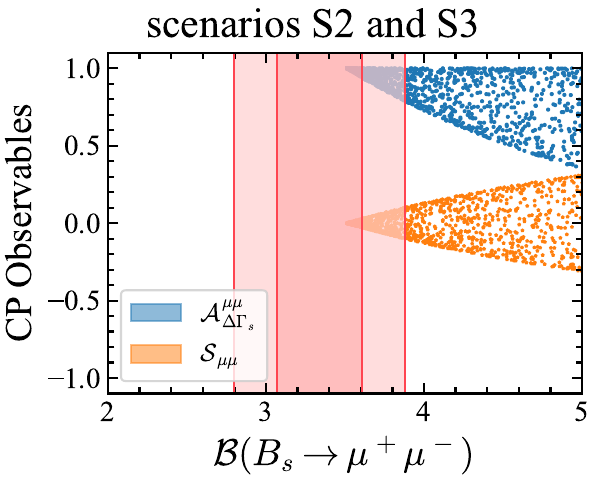}\;
  \includegraphics[width=0.48\linewidth]{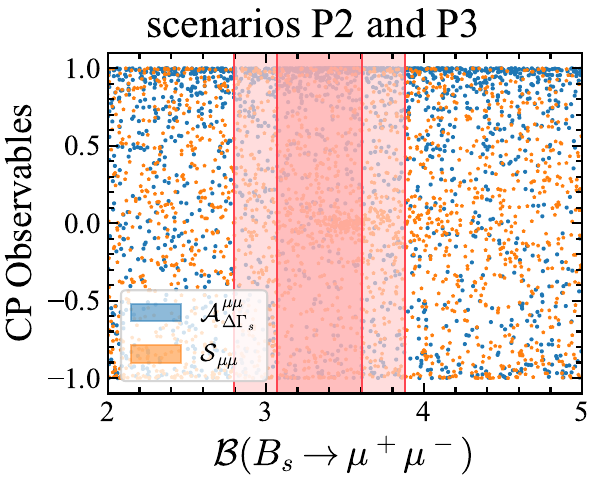}
  \caption{Correlations between the branching ratio and the CP observables of the $\Bsmumu$ decay in the cases of $\ymms \neq 0$ and $\ymmp = 0$ (left), and $\ymms = 0 $ and $\ymmp \neq 0$ (right), with $\theta_L = \theta_R$. The experimental measurement of $\BR$ at the $1\sigma$ (dark red band) and $2\sigma$ CLs (light red band) are also shown.}
  \label{fig:prediction:CPV}
\end{figure}

From the above numerical analysis, we can see that the $\Bsmumu$ decay and the neutron EDM are complementary to each other in probing the CP phases of the $tcS$ couplings. The advantage of the neutron EDM is that it does not depend on the lepton couplings. In figures \ref{fig:complex:LR}\,--\ref{fig:complex:LR:2}, if the coupling $\ymms$ or $\ymmp$ is taken to be smaller than the benchmark values, the constraints from $\BR$ will become looser and could be weaker than the neutron EDM bounds in most of the parameter spaces. The second advantage is that, when the scalar $S$ becomes heavier, the allowed regions by the neutron EDM remain almost unchanged, while the allowed regions by $\BR$ become much larger. This can be understood as follows: the NP contributions to the $\Bsmumu$ decay amplitude and the neutron EDM are both proportional to $m_S^{-2}$; however, the loop functions $f_{1,2}(x_{tS})$ (cf. eqs.~(\ref{eq:f1_function_for_CEDM}) and (\ref{eq:f2_function_for_Weinberg_operator})) compensate for the $m_S^{-2}$ suppression and make the neutron EDM depend weakly on the scalar mass $m_S$, which can also be seen from figure~\ref{fig:EDM}. For the $\Bsmumu$ decay, a key benefit is that the CP observables $\ADG$ and $\Smumu$ depend only on the complex phase $\theta_R$. Therefore, maximum deviations of $\ADG$ and $\Smumu$ from their corresponding SM predictions can be achieved in the limit of $\theta_L = \theta_R$, in which the neutron EDM is, however, not affected by the $tcS$ couplings (cf. eqs.~(\ref{eq:WC:CEDM:mt}) and (\ref{eq:WC:Weinberg:mt}) together with the relation $\operatorname{Re}(\ycts \yctp^*)= \operatorname{Im}(\yctr \yctl^*)/2$). In this limit, we show in figure \ref{fig:prediction:CPV} the correlations between the branching ratio and the CP observables of the $\Bsmumu$ decay for $\ymms \neq 0$ or $\ymmp \neq 0$. In this figure, all the parameter points satisfy the constraints from $(g-2)_\mu$ and lie below the expected $95\%$ CL upper bounds at the LHC for $R < 10$. In the case of $\ymms \neq 0$ and $\ymmp = 0$ (including both scenarios \scenario{S}{2} and \scenario{S}{3}), the two CP observables are close to their SM values, namely in the ranges of $0.8 < \ADG < 1.0$ and $-0.1 < \Smumu < 0.1$ after considering the constraints from $\BR$. However, in the case of $\ymms = 0$ and $\ymmp \neq 0$ (including both scenarios \scenario{P}{2} and \scenario{P}{3}), both $\ADG$ and $\Smumu$ can largely deviate from their corresponding SM expectations. In particular, we find that large deviations always correspond to large NP amplitudes, which have even larger magnitudes but different complex phases compared to the SM contribution. Interestingly, such large NP contributions in the limit of $\theta_L = \theta_R$ survive the constraints from all the observables including the neutron EDM, which can provide constraints only in the general case of $\theta_L \neq \theta_R$.

\section{Conclusions}
\label{sec:conclusions}

It is known that new scalar singlets under the SM gauge group appear naturally in many well-motivated NP scenarios. In this paper, starting from an EFT framework, we have systematically investigated the effects of a new light scalar singlet $S$ with top-quark FCNC interactions. Its effects on the $\Bsmumu$ decay, $(g-2)_\mu$ and the neutron EDM are studied in detail. The scalar singlet $S$ can also induce the rare top-quark decay $t \to cS$ and the single-top production associated with an $S$ at the LHC. Considering the decays of the scalar singlet to a pair of $b$ quarks and/or muons, we have also performed a detailed collider simulation to explore the sensitivity at the current and future LHC to such a scalar singlet $S$ with top-quark FCNC couplings.

It is found that the $\Bsmumu$ decay involves the same scalar couplings as the LHC direct searches of the scalar $S$ in the three-lepton channel, and can provide an efficient probe of such a scalar singlet. For the $\Bsmumu$ decay, the scalar singlet $S$ induces two effective operators $\mO_{S,P}$ at the $b$-quark mass scale $\mu_b$. Their Wilson coefficients $\mC_{S,P}^\NP$ are proportional to the CKM factor $V_{cs}^*V_{tb}$, being much larger than $V_{ts}^*V_{tb}$ associated with the dominant SM contribution $\mC_{10}^\SM$. Furthermore, the scalar contribution does not suffer from helicity suppression. As a result, the branching ratio of the $\Bsmumu$ decay provides a strong constraint on the scalar couplings. After considering the constraints from $(g-2)_\mu$, the bounds on $|\yctr|$ from $\BR$ are stronger than (compatible with) the sensitivity at the LHC with $150\fbi$ for $R=y_{S,P}^{bb}/y_{S,P}^{\mu\mu}\approx 10$ in the case that CP is violated (conserving) in the lepton sector.

It should be noted that the $tcS$ couplings $y_{L,R}^{ct}$ (or $y_{S,P}^{ct}$) can generally carry arbitrary CP-violating phases, which are different from the flavour-conserving $\mu\mu S$ and $bb S$ couplings that must be real as required by hermiticity. In this case, the CP observables $\ADG$ and $\Smumu$ of the $\Bsmumu$ decay provide a useful probe of the CP-violating $tcS$ interactions. We also found that the neutron EDM are very complementary for exploring such CP-violating sources in most of the parameter spaces. However, the neutron EDM depends only on the phase difference $\theta_L-\theta_R$ with $\theta_{L,R}$ denoting the phase of $y_{L,R}^{ct}$, while the $\Bsmumu$ decay depends only on the phase $\theta_R$. In the special limit of $\theta_L=\theta_R$, the neutron EDM is not affected by the $tcS$ interactions, and only the $\Bsmumu$ decay can access the CP violation of the $tcS$ couplings. In this limit, maximum deviations of $\ADG$ and $\Smumu$ from their corresponding SM predictions can be achieved. After considering other relevant constraints, the CP observables can lie in the ranges of $ \ADG \in (0.8,\, 1.0)$ and $\Smumu \in (-0.1,\, 0.1)$ in the case of $\ymms \neq 0$ and $\ymmp=0$ (including scenarios \scenario{S}{2} and \scenario{S}{3}). On the other hand, the two CP observables can still deviate largely from their SM expectations in the case of $\ymms = 0$ and $\ymmp \neq 0$ (including scenarios \scenario{P}{2} and \scenario{P}{3}), namely $ \ADG \in (-1.0,\, 1.0)$ and $\Smumu \in (-1.0,\, 1.0)$.

The results obtained in this paper can be applied to the cases with a scalar singlet $S$ being heavier than around $10\GeV$. For a scalar singlet $S$ with lower mass, its degree of freedom cannot be integrated out and has to be kept in the low-energy effective Lagrangian when describing the low-energy processes like the $\Bsmumu$ decay. Furthermore, a light scalar singlet $S$ can even be produced on shell in $B$-meson decays, and displaced $S$ decays are then needed to be considered. We leave a detailed study of these aspects for future works.

\acknowledgments
This work is supported by the National Natural Science Foundation of China under Grant Nos.~12475094, 12135006, and 12075097, as well as by the self-determined research funds of CCNU from the colleges’ basic research and operation of MOE under Grant Nos.~CCNU24AI003, CCNU22LJ004 and CCNU19TD012. XY is also supported in part by the Startup Research Funding from CCNU. XY thanks Wei Chao and Jinlin Fu for useful discussions.

\appendix

\section{Combined analysis in the scalar and pseudoscalar basis}
\label{sec:combined analysis:SP basis}

\begin{figure}[t]
  \centering
  \includegraphics[width=\linewidth]{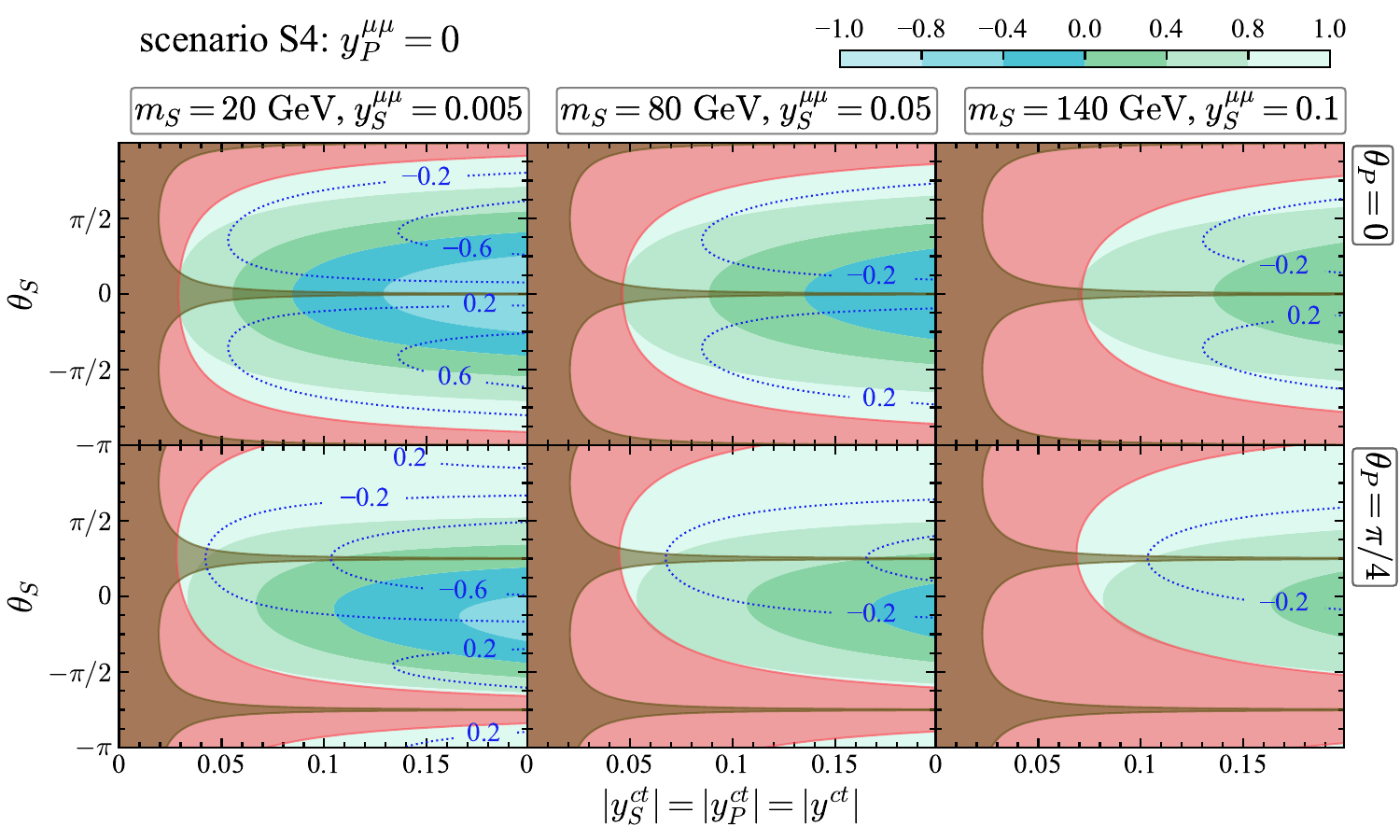}
  \vspace{0.5em}
  \includegraphics[width=\linewidth]{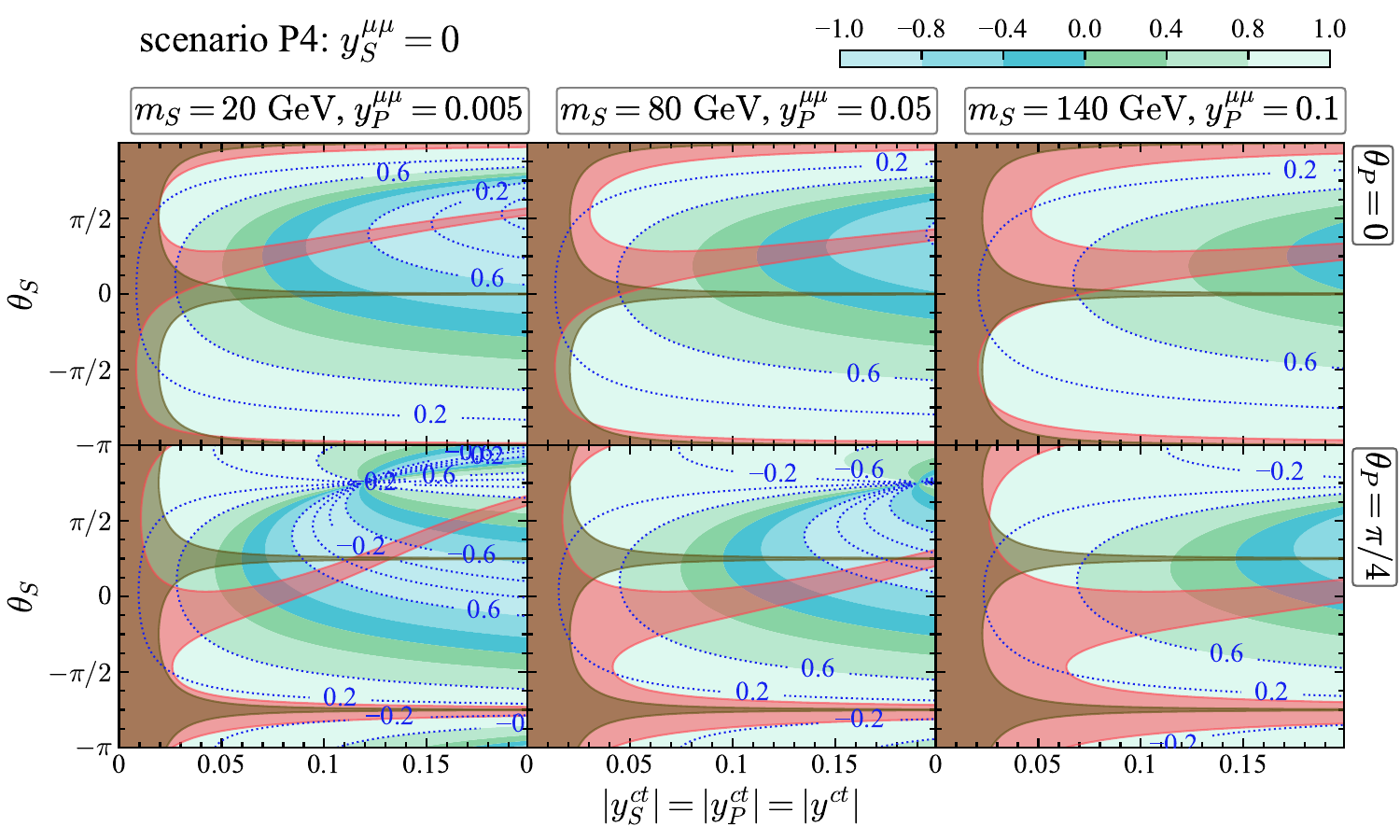}
  \caption{Constraints on $(|y^{ct}|,\, \theta_S)$ in scenarios \scenario{S}{4} (top) and \scenario{P}{4} (bottom). In the second row of each panel, results of $\theta_P=\pi /4$ are shown to illustrate the effects of a nonzero $\theta_P$. The other captions are the same as in figure~\ref{fig:complex:LR}.}
  \label{fig:complex:SP}
\end{figure}

In this appendix, a combined analysis of the CP-violating $tcS$ couplings is performed in the basis of $\ycts$ and $\yctp$, instead of $\yctl$ and $\yctr$ used in subsection~\ref{sec:combined analysis:CPV}. In this basis, CP violation originates from the nonzero phases of $\ycts$ and $i\yctp$, i.e., $\theta_S \neq 0$ and/or $\theta_P \neq 0$, with the definitions $\ycts=|\ycts|e^{i\theta_S}$ and $i\yctp=|\yctp|e^{i\theta_P}$. In this case, the relevant free parameters involve $|\ycts|$, $\theta_S$, $|\yctp|$, $\theta_P$, $\ymms$, and $\ymmp$. In the numerical analysis, we consider the following four scenarios:
\begin{align}
  \big( |\ycts|,\,  \theta_S,\, |\yctp|,\, \ymms,\, \ymmp \big)=
  \begin{cases}
    \big(  |y^{ct}|        ,\;\;    \times  ,\, |y^{ct}| ,\;\; v  ,\;\; 0  \big) & \text{scenario \scenario{S}{4},} \\
    \big(  \;\;\,\,\times ,\;\;\; v       ,\, \;\;\,\,\times ,\;\; v  ,\;\;0  \big) & \text{scenario \scenario{S}{5},} \\
    \big(  |y^{ct}|        ,\;\;    \times  ,\, |y^{ct}| ,\;\; 0  ,\;\; v  \big) & \text{scenario \scenario{P}{4},} \\
    \big(  \;\;\,\,\times ,\;\;\; v       ,\, \;\;\,\,\times ,\;\; 0  ,\;\;v  \big) & \text{scenario \scenario{P}{5},} \\
  \end{cases}
\end{align}
where $\theta_P=0$ is assumed, since only the relative phase between $\ycts$ and $i\yctp$ enters the CP-violating observables.\footnote{In order to illustrate the effects of a nonzero $\theta_P$, results of $\theta_P= \pi /4$ are shown in figure~\ref{fig:complex:SP}.} In the lepton sector, CP is conserved in scenarios \scenario{S}{4} and \scenario{S}{5}, but violated in scenarios \scenario{P}{4} and \scenario{P}{5}.

\begin{figure}[t]
  \centering
  \includegraphics[width=\linewidth]{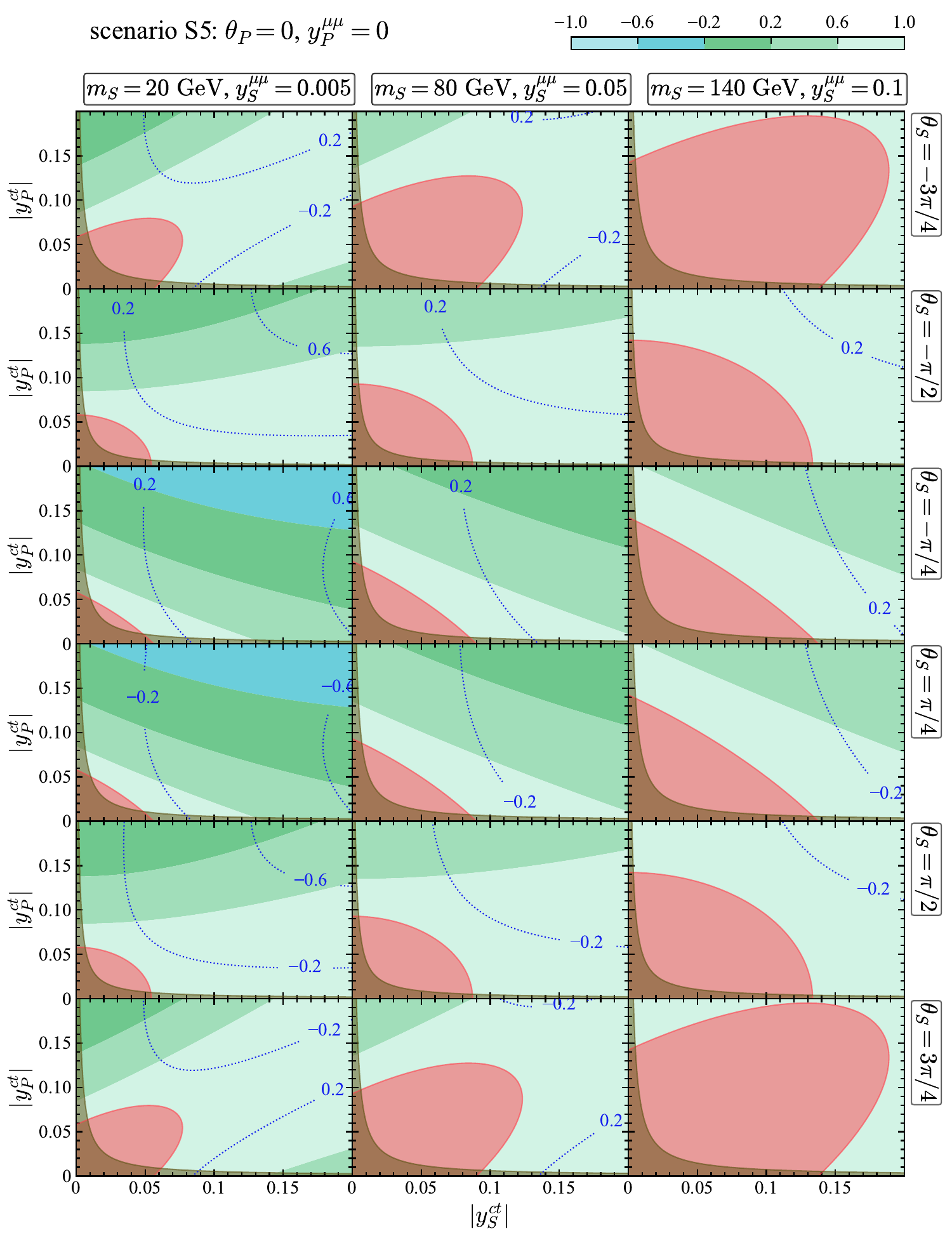}
  \caption{Constraints on $(|\ycts|,\, |\yctp|)$ in scenario \scenario{S}{5} for $\theta_S= \pm\pi/4$, $\pm\pi/2$, $\pm 3\pi/4$. Here $\ymms=0.005$ (left), $0.05$ (middle) and $0.1$ (right) are taken for $m_S=20$, $80$ and $140\GeV$, respectively. The other captions are the same as in figure~\ref{fig:complex:LR}.}
  \label{fig:complex:SP:gSxgP1}
\end{figure}

\begin{figure}[t]
  \centering
  \includegraphics[width=\linewidth]{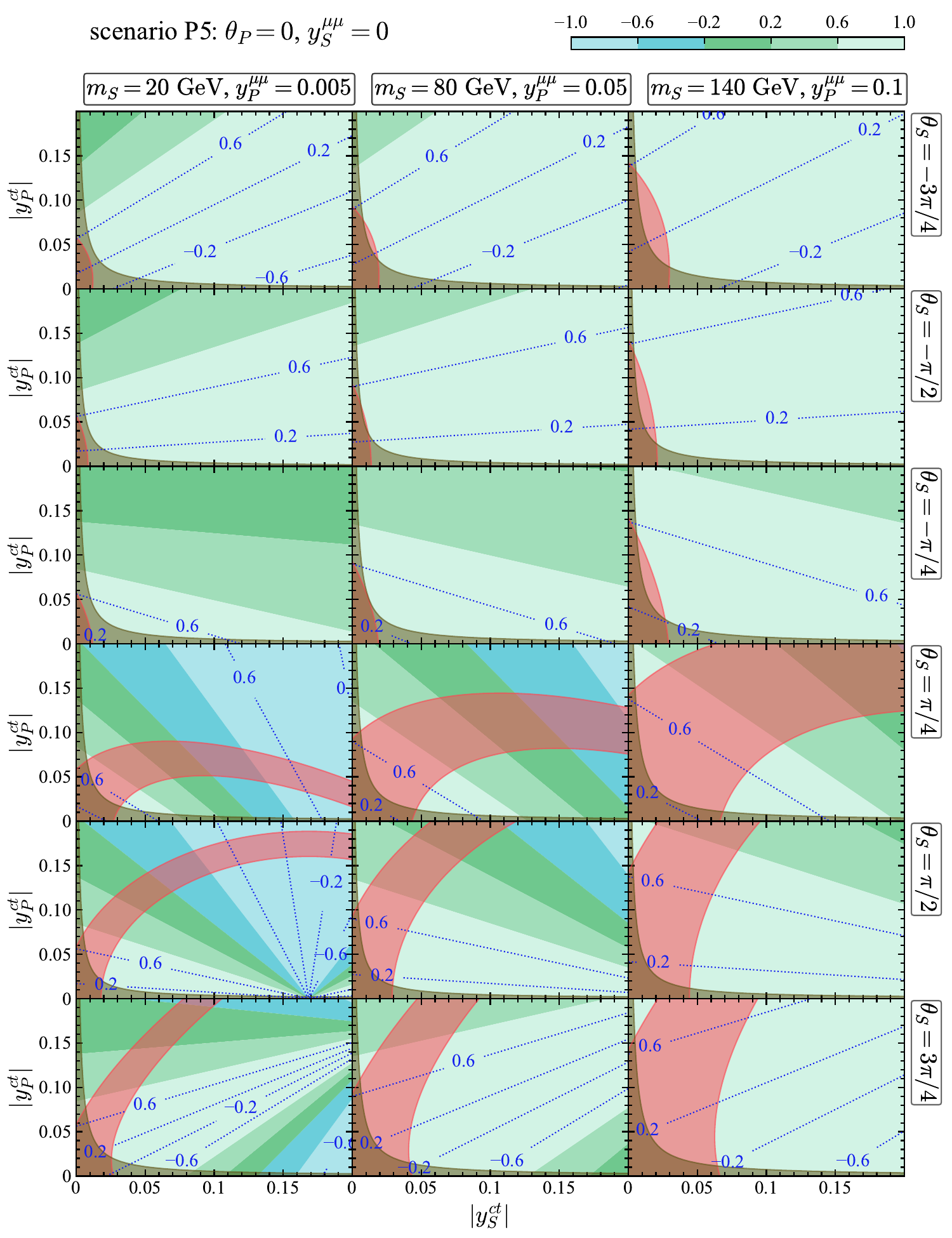}
  \caption{Constraints on $(|\ycts|,\, |\yctp|)$ in scenario \scenario{P}{5} for $\theta_S= \pm\pi/4$, $\pm\pi/2$, $\pm 3\pi/4$. Here $\ymmp=0.005$ (left), $0.05$ (middle) and $0.1$ (right) are taken for $m_S=20$, $80$ and $140\GeV$, respectively. The other captions are the same as in figure~\ref{fig:complex:LR}.}
  \label{fig:complex:SP:gSxgP2}
\end{figure}

In scenarios \scenario{S}{4} and \scenario{P}{4}, we have $|\ycts| = |\yctp| = |y^{ct}|$. Constraints on the parameters from the branching ratio $\BR$ and the neutron EDM are shown in figure~\ref{fig:complex:SP}. Furthermore, the predicted CP observables $\ADG$ and $\Smumu$ of the $\Bsmumu$ decay are also given in figure~\ref{fig:complex:SP}. From this figure, the following observations are made:
\begin{itemize}
\item In the region far from $\theta_S=\pm\pi$, after combining the constraints from $\BR$ and the neutron EDM, the parameter regions with large $|y^{ct}|$ are already excluded. In particular, the regions with large $|y^{ct}|$ allowed by $\BR$ are excluded by the neutron EDM in scenario \scenario{P}{4}.

\item In the region near $\theta_S = \pm \pi$, the constraints from the neutron EDM are stronger than from $\BR$. However, when the values of $\ymms$ and $\ymmp$ are larger than the ones used in figure~\ref{fig:complex:SP}, the constraints from $\BR$ become more stringent and can even be stronger than the neutron EDM bounds.

\item The two CP observables $\ADG$ and $\Smumu$ cannot deviate from their corresponding SM predictions by around $\pm 0.2$, after considering the constraint from $\BR$ and the neutron EDM.
\end{itemize}

In scenarios \scenario{S}{5} and \scenario{P}{5} (i.e., $|\ycts| \neq |\yctp|$), we consider the benchmark phases $\theta_S=\pm\pi/4$, $\pm\pi/2$ and $\pm 3\pi/4$,\footnote{The cases of $\theta_S=0$ and $\pm\pi$ respect the CP symmetry in the quark sector and have already been discussed in subsection~\ref{sec:combined analysis:CPC}.} and show the constraints on the parameter space in figures~\ref{fig:complex:SP:gSxgP1} and \ref{fig:complex:SP:gSxgP2}, respectively. It can be seen that, in the regions with $|\ycts|\gg|\yctp|$ or $|\ycts|\ll|\yctp|$, the branching ratio $\BR$ provides always stronger constraints, even when larger values of $y_{S,P}^{\mu\mu}$ than those used in figures \ref{fig:complex:SP:gSxgP1} and \ref{fig:complex:SP:gSxgP2} are considered. In the region with $|\ycts| \sim |\yctp|$, the neutron EDM puts more stringent bounds. Especially, as shown in figure~\ref{fig:complex:SP:gSxgP2}, the possibility of large NP effects on the $\Bsmumu$ amplitude is excluded by the neutron EDM. In figures~\ref{fig:complex:SP:gSxgP1} and \ref{fig:complex:SP:gSxgP2}, predictions on the CP observables $\ADG$ and $\Smumu$ are also shown. We can see that, different from scenario \scenario{P}{5}, the cases of $\pm\theta_S$ in scenario \scenario{S}{5} (cf. figure~\ref{fig:complex:SP:gSxgP1}) cannot be distinguished by $\BR$, $\ADG$ or the neutron EDM. However, the observable $\Smumu$ allows us to resolve this twofold ambiguity, since its sign will be  different for $\theta_S$ and $-\theta_S$.

Since the couplings $\ycts$ and $\yctp$ are equivalent to $\yctl$ and $\yctr$ for describing the $tcS$ interactions, predictions on the CP observables are the same in these two bases, which are already given in figure \ref{fig:prediction:CPV}.

  
\bibliographystyle{JHEP}
\bibliography{refs.bib}

\end{document}